\documentclass[preprint2,natbib209]{aastex}
\usepackage{graphicx}
\usepackage{amsmath}
\usepackage{amssymb}
\usepackage{natbib}
\usepackage{cases}
\usepackage{paralist}
\usepackage{bm}
\usepackage{upgreek}
\usepackage{bbold}
\usepackage{url}
\usepackage{blindtext}
\newcommand{\taurex}{$\mathcal{T}$-REx}

\DeclareMathAlphabet{\mathpzc}{OT1}{pzc}{m}{it}

\shorttitle{TauREx}
\shortauthors{Waldmann et al.}

\begin{document}

\title{Tau-REx II: Retrieval of emission spectra}

\author{I. P. Waldmann, M. Rocchetto, G. Tinetti, E.J. Barton, S.N. Yurchenko,
J. Tennyson}
\affil{Department of Physics \& Astronomy, University College London, Gower
Street, WC1E 6BT, UK}
\email{ingo@star.ucl.ac.uk}

\begin{abstract}

 \taurex~(Tau Retrieval of Exoplanets) is a novel, fully Bayesian atmospheric
retrieval code custom built for extrasolar atmospheres. In \citet{waldmann15}
the transmission spectroscopic case was introduced, here we present the emission
spectroscopy spectral retrieval for the \taurex\ framework. Compared to
transmission spectroscopy, the emission case is often significantly more
degenerate due to the need to retrieve the full atmospheric
temperature-pressure (TP) profile. This is particularly true in the case of
current measurements of exoplanetary atmospheres, which are either of low
signal-to-noise, low spectral resolution or both. Here we present a new
way of combining two existing approaches to the modelling of the said TP profile: 1)
the parametric profile, where the atmospheric TP structure is analytically
approximated by a few model parameters, 2) the Layer-by-Layer approach, where
individual atmospheric layers are modelled. Both these approaches have distinct
advantages and disadvantages in terms of convergence properties and potential
model biases. The \taurex~hybrid model presented here is  a new two-stage TP
profile retrieval, which combines the robustness of the analytic solution with
the accuracy of the Layer-by-Layer approach. The retrieval
process is demonstrated using simulations of the hot-Jupiter WASP-76b and the hot SuperEarth
55~Cnc~e, as well as on the secondary eclipse measurements of HD189733b.

\end{abstract}

\keywords{methods: data analysis --- methods: statistical  --- techniques:
spectroscopic --- radiative transfer }

\section{Introduction}

The characterisation of extrasolar planets through the spectroscopic
measurements of their atmospheres has become a well established field today \citep{2013A&ARv..21...63T}. 
In \citet{waldmann15}, we presented the \taurex~(Tau Retrieval of Exoplanets)
suite for transmission spectroscopic measurements. In this paper we introduce
the corresponding retrieval for emission spectroscopic data. 

Transmission and emission spectroscopy carry highly complementary aspects.
Whereas transmission spectroscopy is less sensitive to thermal gradients, the
emission spectroscopy case probes the full temperature-pressure profile
(TP-profile hereafter) of the atmosphere. This makes the emission case
significantly harder to constrain without the luxury of in-situ measurements.
\citet{1958sues.conf..133K} was the first to suggest remote sensing of the planetary
atmospheric temperature structure through the infra-red, thermal radiance of the
planet. \citet{1959JOSA...49.1004K} expanded on this pioneering work by laying
the foundations of retrieving exact TP-profiles from emission spectroscopic
measurements. Remote sensing of planetary atmospheres in our solar system has
been a long story of success 
\citep[e.g.][]{1969Sci...165.1256W,Conrath:1970we, 1972Sci...175..305H,1973JGR....78.4267C,1976RvGSP..14..609R,1981Sci...212..192H,2007Icar..189..457F,2008JQSRT.109.1136I} 
as well as \citet{Hanel:2003hq}, and
references therein. More recently, emission spectroscopy remote sensing has been
expanded to exoplanetary atmospheres
\citep{2009ApJ...707...24M,Lee:2011gl,2012ApJ...749...93L,2014RSPTA.37230086G}.
\citet{2013ApJ...775..137L} provides a  comparison of existing exoplanetary
emission spectroscopy retrieval codes.

\subsection{\taurex}

\taurex~ is a novel, Bayesian atmospheric spectroscopy retrieval suite designed
for extrasolar planets. In \citet{waldmann15}, hereafter W15, we introduce the
overall architecture, data handling, minimisation/sampling routines, handling of
molecular line lists, Bayesian model selection and the transmission spectroscopy
forward model. In this publication we present the emission spectroscopy forward
model as well as the atmospheric temperature-pressure (TP) profile retrieval
implemented. 

\begin{figure}[h]
\centering
\includegraphics[width=\columnwidth]{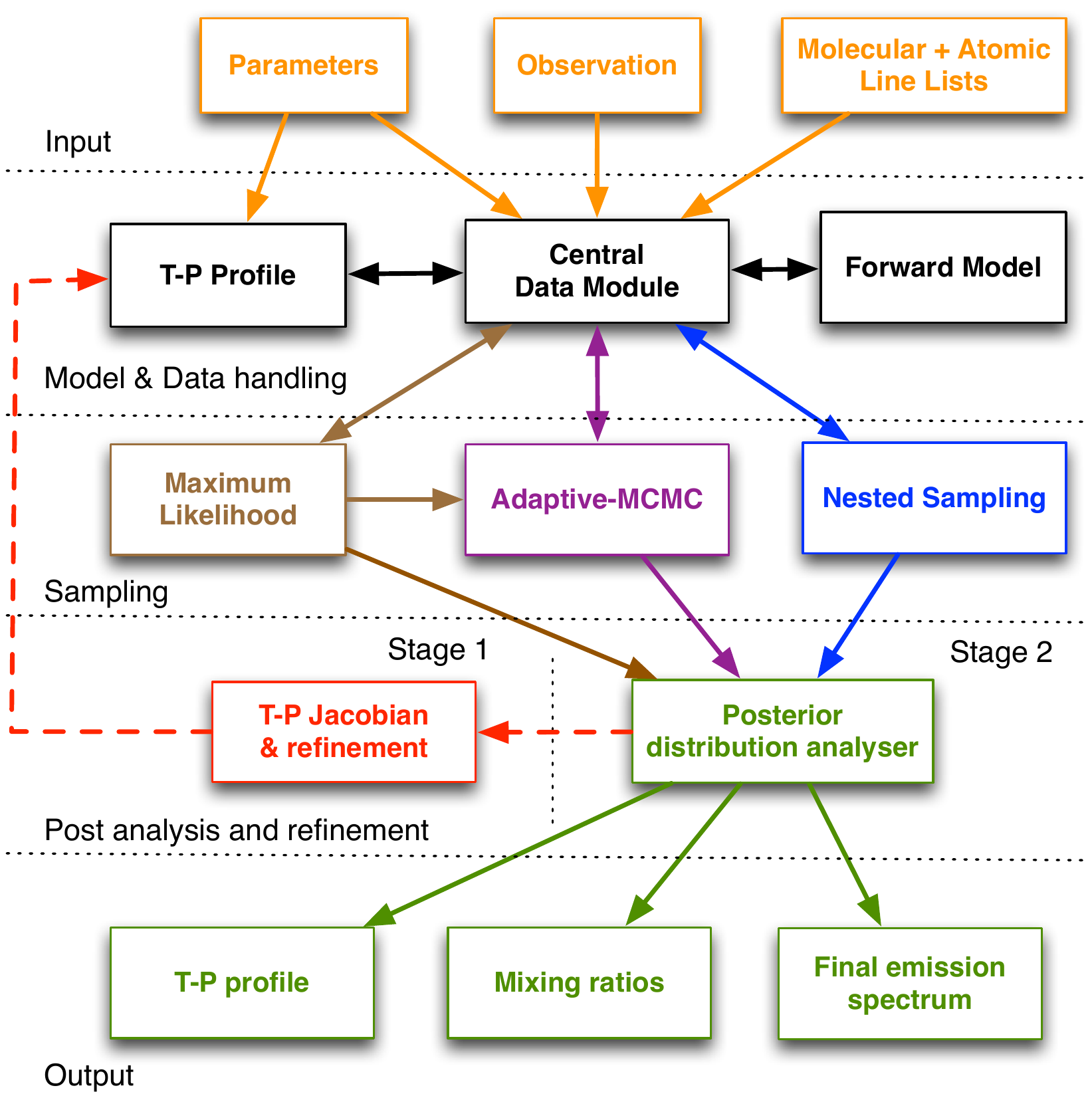}
\caption{Flowchart illustrating the modular design of \taurex. This flowchart is
an update from \citet{waldmann15}, reflecting the differing architecture of the
emission spectroscopy code.  \label{fig:flowchart}}
\end{figure}

Figure~\ref{fig:flowchart} shows a schematic diagram of the \taurex~architecture for the
emission retrieval. The program can be subdivided into five sections: 

\begin{enumerate}
\item {\it Inputs}: Program inputs such as parameter files, molecular line-lists
\citep[ExoMol][]{Tennyson:2012ca}, \citep[HITEMP][]{2010JQSRT.111.2139R},
stellar spectra \citep[PHOENIX,][]{Allard:2012jx} and spectroscopic observations to be
analysed.
\item {\it Model \& Data handling}: The {\it Central Data} Module manages all
calls to the {\it TP-Profile} and {\it Forward Model} Modules and provides a
standardised, common interface to different sampling routines implemented. 
\item{\it Sampling}: \taurex~features three independent sampling routines:
Maximum Likelihood estimation (MLE, based on the LM-BFGS minimisation in W15),
Markov Chain Monte Carlo (MCMC) and Nested Sampling (NS) routines. We refer the
reader to W15 for implementation details. 
\item {\it Post analysis \& refinement}: This stage differs from W15 by
providing a two iteration stages for the determination of the TP-Profile, see
section~\ref{sec:tp}.
\item {\it Output}: The final posterior distributions are analysed and the final
model spectrum, TP-Profile and mixing ratios are returned. 
\end{enumerate}

\section{Forward Model}

We briefly describe the radiative transfer forward model, for a more exhaustive
discussion we refer the reader to the standard literature
\citep[e.g.,][]{chandrasekhar1960,Yung:vp,Liou:2002uh,Hanel:2003hq,Mihalas:2013vm}.
In what follows, we closely follow the nomenclature of \citet{Liou:2002uh}. For
non-scattering atmospheres in local thermodynamic equilibrium, the basic
equation describing the thermal radiation of an atmosphere is given by the
Schwartzschild equation:

\begin{equation}
\mu \frac{d I_{\lambda} (\tau, \mu)}{d\tau} = I_{\lambda} (\tau, \mu) -
B_{\lambda} (T) 
\end{equation}

\noindent where $I_{\lambda}$ is the intensity per wavelength, $\lambda$,
$B_{\lambda}$ is the Planck function at temperature $T$, $\mu =
\text{cos}\theta$ is the upwards inclination, $\tau$ is the overall optical
depth, given as function of altitude ($z$) by 

\begin{equation}
{\tau}_{\lambda}(z) = \sum_{m=1}^{N_{m}} {\tau}_{\lambda,m}(z)
\end{equation}

\noindent where $\tau_{\lambda,m}$ denotes the optical depth per absorbing
species, $m$, given by 
\begin{equation}
\tau_{\lambda,m} = \int_{z}^{z_{\infty}} {\varsigma}_{\lambda,m} (z')
\chi_{m}(z') \rho_{N} (z') \text{d}z'.
\end{equation}

\noindent Here ${\varsigma}_{\lambda,m}  $ is the absorption cross section,
$\chi_{m}$ the column density and $\rho_{N}$ the number density. We can now
express the upwards welling radiance as

\begin{equation}
I_{\lambda}(\tau,\mu) = I_{\lambda}(\tau_{s}) e^{-(\tau_{s}-\tau)/\mu} +
\int_{\tau}^{\tau_{s}} B_{\lambda}(T_{\tau'}) e^{-(\tau' - \tau)/\mu}
\frac{\text{d}\tau'}{\mu}
\end{equation}

\noindent where the first right-hand-side term is the radiation at the planetary
surface (or defined surface pressure for gaseous planets), and the second term
denotes the integrated emission contributions for individual plane-parallel
layers. The monochromatic transmittance and its derivative (weighting function)
can be defined as 

\begin{equation}
\mathcal{T}_{\lambda} (\tau) = e^{-\tau},~~~~\frac{\partial \mathcal{T}_{\lambda} (\tau)}{\partial
\tau} = - e^{-\tau}.
\end{equation}

\noindent Hence we express the total integrated radiation at the top of the
atmosphere (TOA, $\tau = 0, z = \infty$)  as

\begin{equation}
I_{\lambda}(\tau=0) = B_{\lambda}(T_{s}) e^{-\tau_{s}} +
\int_{\tau_{s}}^{0} B_{\lambda}(T_{\tau})\frac{\partial \mathcal{T}_{\lambda} (\tau)}{\partial
\tau} d\tau
\end{equation}

\noindent where $\tau_{s}$ and $T_{s}$ are the optical depth and temperature at
the planetary surface. The final exoplanetary emission spectrum is now given by 

\begin{equation}
F_{P}/F_{\ast} = \frac{I_{\lambda}(\tau=0)}{I_{\ast}} \times \left (
\frac{R_{P}}{R_{\ast}} \right )^{2}
\end{equation}

\noindent here $I_{\ast}$ is the stellar intensity. By default
\taurex~interpolates the stellar flux from a library of {\sc{Phoenix}}\footnote{\url{
http://www.hs.uni-hamburg.de/EN/For/ThA/phoenix/index.html}} stellar spectra, gridded at 100K intervals. Alternatively \taurex~can approximate the
stellar intensity using a black-body at the stellar temperature.

\section{Temperature - Pressure Profile}
\label{sec:tp}

The determination of the vertical atmospheric temperature profile (also referred
to as temperature-pressure profile or TP-profile for short) is one of the key
challenges in the retrieval of atmospheric emission spectra. Typically two
approaches exist in the retrieval of the TP-profile: 1) Layer-by-layer
retrieval, 2) Analytic parameterisation. 

1) The layer-by-layer approach consists of modelling the temperature of each
plane-parallel atmospheric layer independently. This approach is usually adopted
for retrieval work of the Earth's atmosphere and solar system planets
\citep{Hanel:2003hq,Rodgers:2000tw}. The advantage of such an approach is its
non-parametric nature, i.e. no prior knowledge is imposed on the temperature
retrieval nor is the solution limited by potential restrictions and biases of
an analytic TP-profile. The obvious disadvantage lies in the potentially
poor convergence properties of such an approach in low S/N scenarios. Often,
data quality or sparse spectral sampling do not allow for a pure layer-by-layer retrieval due to
the large number of free parameters incurred. This is particularly true for
current exoplanet spectroscopy data, which is either of relatively low S/N, low
spectral resolution, or both. 
A common approach adopted is to impose a `regularisation' of the
TP-profile based on the physical reality that adjacent atmosphere layers should
exhibit some correlation in temperature. 

2) The second approach to the TP-profile retrieval is to adopt an analytic (here
also referred to as `parametric') model. Such models analytically approximate
the mean underlying physics of the atmospheric thermal structure. Such models
have the advantage of containing far fewer free parameters compared to the layer-by-layer approach, hence they converge faster. However, the solution is
constrained within the bounds of the model assumed. 

In summary, the layer-by-layer methodology is most objective but features poor
convergence properties, whilst the parametric model is less objective but
converges more robustly. 
For a review of existing implementations of both approaches in the field of
exoplanet atmospheric retrieval we refer the reader to
\citet{2013ApJ...775..137L}.  

As described in section~\ref{sec:hybrid}. \taurex~employs a hybrid model
combining both, parametric and layer-by-layer approaches in a two stage
retrieval process. In section~\ref{sec:parametric} we briefly describe the
parametric models implemented in \taurex~as part of the \taurex~retrieval
framework.

\subsection{Parametric Models in the literature}
\label{sec:parametric}

Analytical global-average TP-profiles exist in various flavours ranging from
2-stream purely radiative to radiative-convective approximations and 
global circulation models (GCMs).
We refer the reader to the relevant literature for a more in-depth discussion on
the various analytic approaches
\citep[e.g.][]{Liou:2002uh,2003ApJ...594.1011H,2008ApJS..179..484H,2008ApJ...678.1436B,0004-637X-699-1-564, 2009ApJ...707...24M,2010ppc..book.....P,Guillot:2010dd,2012MNRAS.420...20H,2014ApJS..215....4H,Robinson:2012ky}. 
For the Stage~1 approach, \taurex~features a purely radiative
2-stream model as well as a choice of simpler TP-profiles based on purely
geometric considerations (see \ref{sec:othertp}). As described in
section~\ref{sec:hybrid}, the exact form of the TP-profile is less relevant in
our case as the results will be refined in a second stage fitting.  

Following \citet{Guillot:2010dd}, the mean global temperature profile for a
simple radiative downstream-upstream approximation can be expressed as 

\begin{equation}
T^4(\tau) = \frac{3 T^{4}_{int}}{4} \left ( \frac{2}{3} + \tau \right ) +
\frac{3 T^{4}_{irr}}{4}  \xi_{\gamma_{1}} (\tau) 
\label{equ:guillottp}
\end{equation} 

\noindent where $T_{int}$ is the planet internal heat flux, $T_{irr}$ the solar
flux at the top of the atmosphere and

\begin{align}
\xi_{\gamma_{i}} =& \frac{2}{3} + \frac{2}{3\gamma_{i}} \left [ 1 + \left (
\frac{\gamma_{i}\tau}{2} - 1 \right ) e^{-\gamma_{i}\tau} \right ] \\
&+ \frac{2\gamma_{i}}{3} \left ( 1 - \frac{\tau^{2}}{2} \right )
E_{2}(\gamma_{i}\tau) \nonumber
\end{align}

\noindent where $\gamma_{1} = \kappa_{\nu}/\kappa_{IR}$ is the ratio of mean
opacities in the optical ($\kappa_{\nu}$) and infra-red ($\kappa_{IR}$) and
$E_{2}$ is the second-order exponential integral given by 

\begin{equation}
E_{n+1}(z) = \frac{1}{n} [ e^{-z} - zE_{n}(z)].
\end{equation}

\noindent We note that similar parameterisations exist in the literature
\citep[e.g., eq. (18), ][]{Robinson:2012ky, 2010ppc..book.....P}. We also include
the variation by  \citet{2012ApJ...749...93L} and \citet{2014A&A...562A.133P}  
including two optical opacity
sources $\kappa_{\nu_{1}}$ and $\kappa_{\nu_{2}}$ and a weighting factor between
optical opacities (left as free parameter) $\alpha$,

\begin{align}
T^4(\tau) =& \frac{3 T^{4}_{int}}{4} \left ( \frac{2}{3} + \tau \right ) +
\frac{3 T^{4}_{irr}}{4} (1 - \alpha) \xi_{\gamma_{1}} (\tau) \\
&+ \frac{3 T^{4}_{irr}}{4} \alpha \xi_{\gamma_{2}} (\tau). \nonumber
\label{equ:linetp}
\end{align} 

\noindent The temperature as function of opacity $\tau$ can be mapped to a
pressure grid by assuming the following relation

\begin{equation}
\tau = \frac{\kappa_{IR} P}{g}.
\end{equation}

\subsubsection{Other TP-profiles}
\label{sec:othertp}

In addition to the above TP-profiles, we include an isothermal profile as well
as a `3-point' and `4-point' profile. The 3-point profile is purely geometric
and keeps the top of atmosphere temperature, $T_{TOA}$, the tropopause
temperature and pressure, $T_{1}$, $P_{1}$, and the surface (or 10 bar pressure)
temperature $T_{10bar}$ as free variables. The TP-profile is then linearly
interpolated in ln$(P)$.  The 4-point profile adds an extra variable
temperature-pressure point to the profile.

\subsection{The \taurex~Hybrid Model}
\label{sec:hybrid}

\begin{figure}[h]
\centering
\includegraphics[width=\columnwidth]{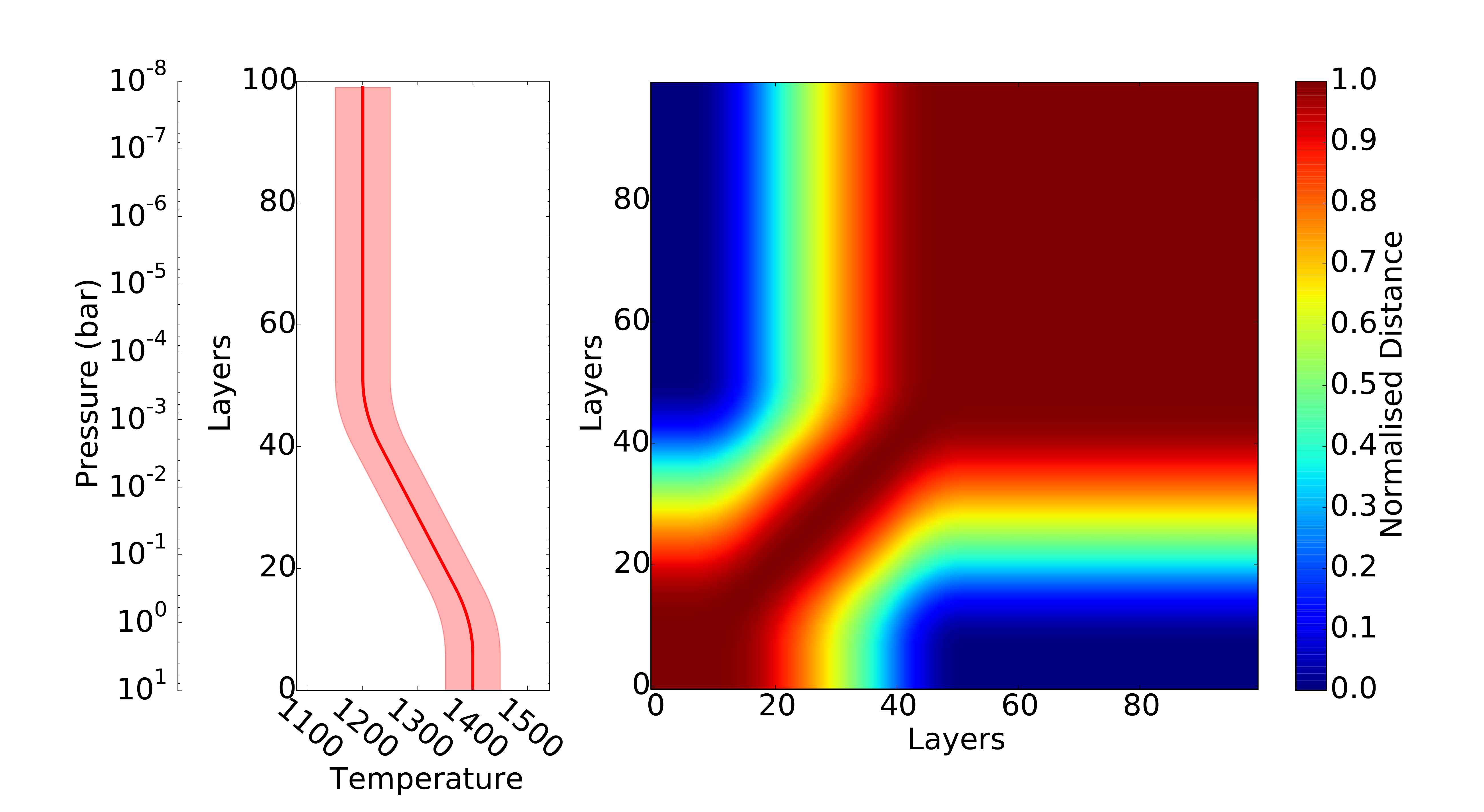}
\caption{Left: A given Temperature - Pressure Profile without inversion. Right:
Correlation matrix $\mathcal{C}$, equation~\ref{equ:tp-matrix2}, of the TP-profile
on the left. Atmospheric layers with similar temperatures feature a high
layer-layer correlation whilst layers with temperature differences feature a
lower correlation.  \label{fig:corrmatrix}}
\end{figure}

In \taurex~we have worked towards a hybrid solution of the above mentioned
approaches in which the final solution is not bound by a parametric model but
does not inherit the potentially poor convergence properties (and susceptibility
to noise) of a layer-by-layer approach. 
Here we proceed in two stages: 1) We perform a parametric model retrieval, 2) we
take the retrieval solution to `guide' a second layer-by-layer retrieval,
relaxing the parametric model constraint and thereby fine-tuning the
TP-profile. 

\subsubsection{Stage 1}

In {\it Stage~1} we adopt a classical parametric model retrieval using one of
the TP-profile parameterisations implemented in \taurex.  The idea is to constrain
an initial best-fit of the model and whilst a realistic model (i.e. one well suited to describe the underlying TP-profile) is preferable, it
is not a hard requirement. We now fit the solution using either of
\taurex's sampling routines (MLE, MCMC, NS). The error on the sampled parametric
model parameters is converted to one $\sigma$ lower and upper temperature bounds for a
layer-by-layer model. This is done using a numerical model permutation analysis of the \emph{Stage~1} parameters. 

We then calculate the following distance matrix 

\begin{equation}
\label{equ:tp-matrix}
\Delta^{2}_{ij} = |\hat{T}_{i} - \hat{T}_{j}|^{2} + (\sigma_{i}+ \sigma_{j})^{2}
\end{equation}

\noindent where $\hat{T}$ is the maximum likelihood temperature estimator of the
parametric model fit for the $i^{th}$ and $j^{th}$ atmospheric layer and
$\sigma$ is the respective error. Equation~\ref{equ:tp-matrix} captures the
layer to layer temperature variations in the TP-profile and is hence
conceptually similar to the Jacobian of the profile. We normalise $\Delta_{ij}$
in terms of the minimal and maximal temperature variations found in the
TP-profile, 

\begin{equation}
\label{equ:tp-matrix2}
\mathcal{C}_{ij} = 1.0 - \frac{\Delta_{ij} -
\text{argmin}~\Delta}{\text{argmax}~\Delta}
\end{equation}

\noindent which can be understood as a positively
defined temperature correlation matrix with layers most similar in temperature
featuring the highest correlation and layers most dissimilar resulting in a very
low correlation. An example of $\mathcal{C}$ for a given TP-profile can be found
in fig.~\ref{fig:corrmatrix}.

\subsubsection{Stage 2}

\begin{figure*}
\centering
\includegraphics[width=\textwidth]{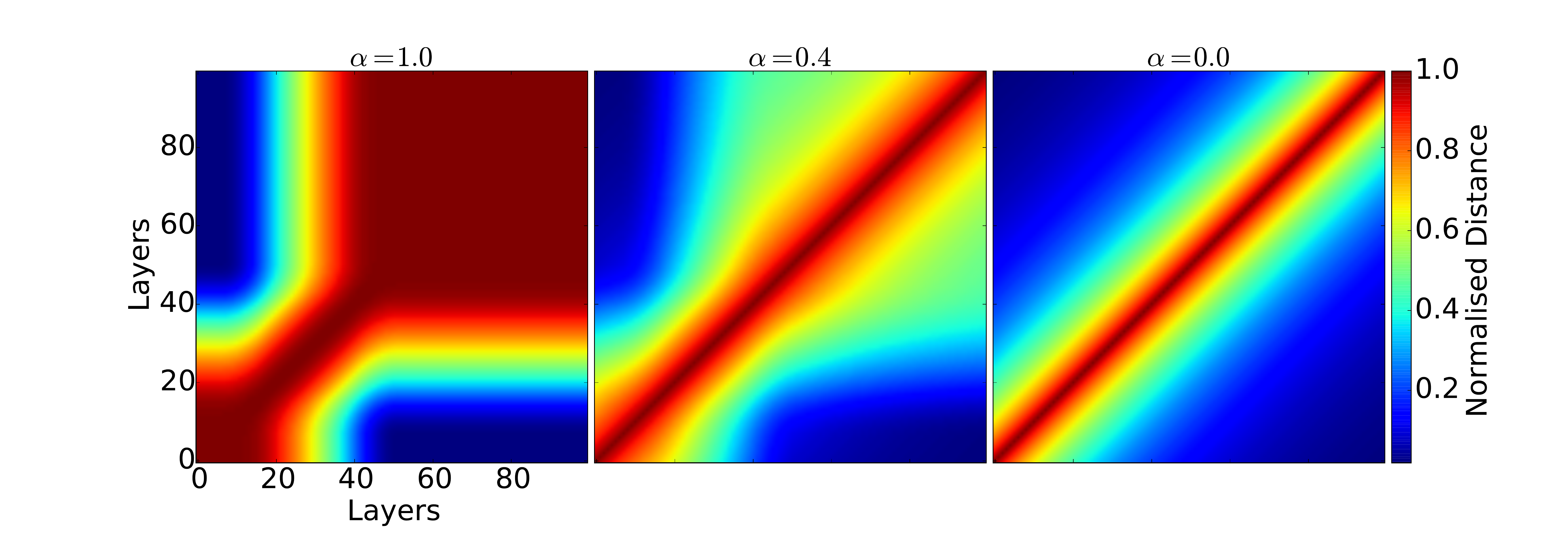}
\caption{Hybrid correlation matrix $\mathcal{D}$,
equation~\ref{equ:hybrid}, at different values of $\alpha$. The left most is the
pure {\it Stage 1} correlation matrix, $\mathcal{C}$, whereas the right plot is
the pure correlation-length-only matrix $S$, eq.~(\ref{equ:rodgers}).
Over-plotted are contours of $\mathcal{D}(\alpha)$. The middle panel shows an intermediate state of left (60$\%$) and right (40$\%$).   \label{fig:hybridmatrix}}
\end{figure*}

In the second stage of our TP-profile retrieval we relax the original solution
found in {\it Stage 1} and by that `fine-tune' the TP-profile. We construct a
second correlation matrix imposing an exponential correlation length across
pressure levels \citep{1976RvGSP..14..609R,Rodgers:2000tw}

\begin{equation}
\label{equ:rodgers}
S_{ij} = (S_{ii}S_{jj})^{1/2} \text{exp} \left (-\frac{| ln (P_{i}/P_{j})| }{c}
\right )
\end{equation}

\noindent where $c$ is the correlation length in terms of atmospheric scale
heights. We set $c = 3.0$  by default, but can otherwise be user defined. \citet{Rodgers:2000tw} gives an advisable range between 1.0 - 8.0, with \citet{2008JQSRT.109.1136I} using a default of $c = 1.5$ and \citet{2013ApJ...775..137L} $c = 7.0$. Larger values of $c$ correspond to a stronger regularisation of the TP-profile (i.e. smoothing). A stronger regularisation can be useful in poorly constrained data sets (either low S/N, sparsely sampled or both). 
Equation~\ref{equ:rodgers} is identical to that used in the layer-by-layer
approach by \citet{2008JQSRT.109.1136I}. We now construct a hybrid correlation matrix by
combining equations~\ref{equ:tp-matrix2}~\&~\ref{equ:rodgers}

\begin{equation}
\label{equ:hybrid}
\mathcal{D}_{ij}(\alpha) =  \alpha \mathcal{C}_{ij} + (1-\alpha) S_{ij}
\end{equation}

\noindent where $\alpha$ is a scaling factor ranging from zero to one.
Figure~\ref{fig:hybridmatrix} shows instances of $\mathcal{D}(\alpha)$ at
varying values of $\alpha$. We now correlate our layer-by-layer TP-profile with
$\mathcal{D}(\alpha)$ whereby leaving $\alpha$ as free parameter to be fitted.
By allowing $\alpha$ to vary, we can dynamically relax the parametric model
solution from {\it Stage 1}, $\alpha = 1.0$, to an unconstrained solution,
$\alpha = 0.0$. The advantage of such an approach is twofold: 1) Since the
{\it Stage 1} fitting should have already achieved a solution close to the real
maximum likelihood, convergence in the second stage towards the volume of lowest
$\chi^{2}$ is significantly faster; 2) \taurex~can freely decide to relax the 
{\it Stage 1} solution should this be favoured by the data. Practically this
happens frequently at later stages of the fitting. Whereas \taurex~initially
converges towards the {\it Stage 1} solution (i.e. $\alpha \rightarrow 1$),  at
later stages of the fitting the code begins to reject the parametric model (i.e.
$\alpha \rightarrow 0$) as it `fine-tunes' the original solution. This `tuning'
can achieve significantly higher maximum likelihoods than the {\it Stage 1}
fitting alone. 

\subsection{Non-linear Sampling}

\begin{figure}
\centering
\includegraphics[width=\columnwidth]{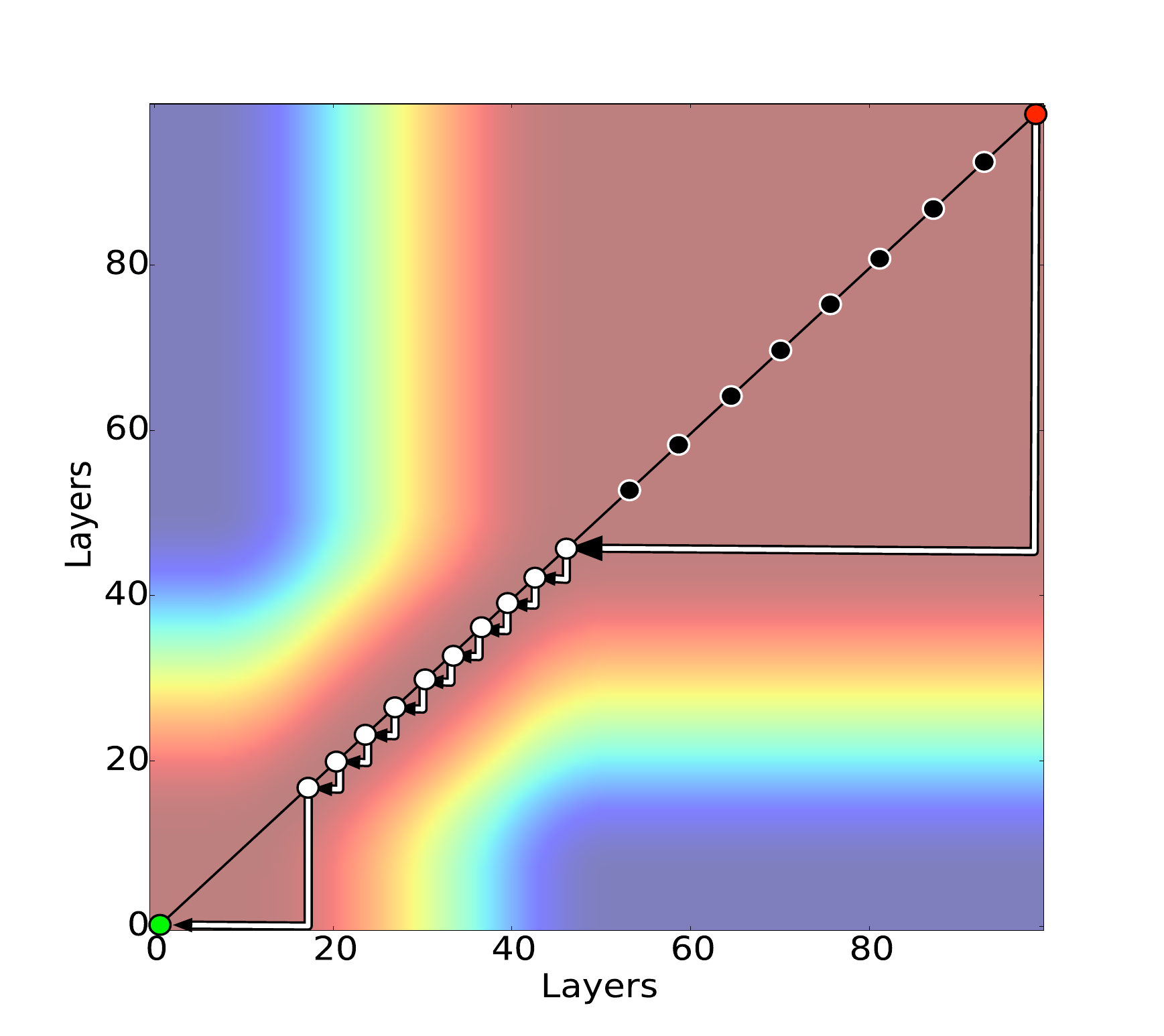}
\caption{Nonlinear TP-profile sampling on correlation matrix
$\mathcal{C}$ (same as right hand plot in figure~\ref{fig:corrmatrix}). Starting
at the top of the atmosphere (red dot) we retain all layers in the TP-profile
that correspond to a change in gradient $> 2\%$ with respect to the previously
retained layer until the bottom layer (green dot) is reached. The selected
layers are denoted by white dots and arrows represent the path of the
compression algorithm across the correlation matrix $\mathcal{C}$. Should no
gradient change be detected for $> 10$ layers, an extra sampling point is
introduced (black dots).  \label{fig:nonlinear}}
\end{figure}

In the approach explained in section~\ref{sec:hybrid}, we keep an even sampling
of atmospheric layers in log($P$) for {\it Stage~2}. For well sampled and high
S/N data, this approach is adequate. However for coarsely sampled and/or poor
S/N data, it is often advisable to reduce the number of free parameters to a
minimum to aid convergence. In these cases we can utilise the sparsity of the
Stage 1 TP-profile solution to devise a nonlinearly sampling of the exoplanetary
atmosphere. We base our compression algorithm on the fact that only changes in
the temperature-pressure gradient need to be modelled, i.e., an isothermal
TP-profile can be perfectly described by a single temperature parameter whereas
areas of non-isothermal structure need more parameters to capture variation. The
compression algorithm uses the {\it Stage~1} correlation matrix $\mathcal{C}$ to
only retain TP-profile layers corresponding to a $> 2\%$ change in the TP
gradient with respect to the previously retained layer. We graphically depict
this in figure~\ref{fig:nonlinear}. In addition to TP-profile layers sampled
this way, we also include an extra sampling layer whenever no change in thermal gradient
was detected for $> 10$ layers. The majority of the TP-profile variations should be captured by the Stage~1 retrieval and Stage~2 is `fine-tuning' this solution. The inclusion of a coarse sampling 
in isothermal regimes does allows the Stage~2 retrieval to deviate from any Stage~1 solution, should this be supported by a higher global likelihood.
Using the non-linear sampling, we can reduce a 100 layer atmospheric
model to typically 15-25 free parameters.

\section{Validation of \taurex}
\label{sec:example}

We demonstrate the emission spectroscopy retrieval with two simulations and the secondary eclipse spectrum of HD189733b. The simulations are as follows: 1) high
S/N observation simulation of a hot Jupiter similar to WASP-76b; 2) observations
of the hot SuperEarth 55~Cnc~e, simulated for a 1 meter class spectroscopic space mission.
In our simulations we opt for two oxidised atmospheres at high temperatures ($> 1500$K).

For each retrieval stage we calculate the global Bayesian Evidence of the
solution set. Here the Bayesian Evidence (E, or simply Evidence hereafter) is
given as the integral of the product of the global likelihood and the prior
space

\begin{equation}
\label{equ:bayesevidence}
E = \int P({\bm \theta} | \mathcal{M}) P({\bf x} | {\bm \theta}, \mathcal{M})
\text{d}{\bm \theta}
\end{equation}

\noindent where $P({\bm \theta} | \mathcal{M})$ is the prior over the parameter
space ${\bm \theta}$, for retrieval model $\mathcal{M}$. $P({\bf x} | {\bm
\theta})$ is the likelihood for the observed data vector ${\bf x}$ given the
parameter space and retrieval model. Retrieval Evidences are reported in
tables~\ref{tb:wasp76} and \ref{tb:55cnc}.

\subsection{WASP-76b}
\label{sec:wasp}

\begin{figure}
\centering
\includegraphics[width=\columnwidth]{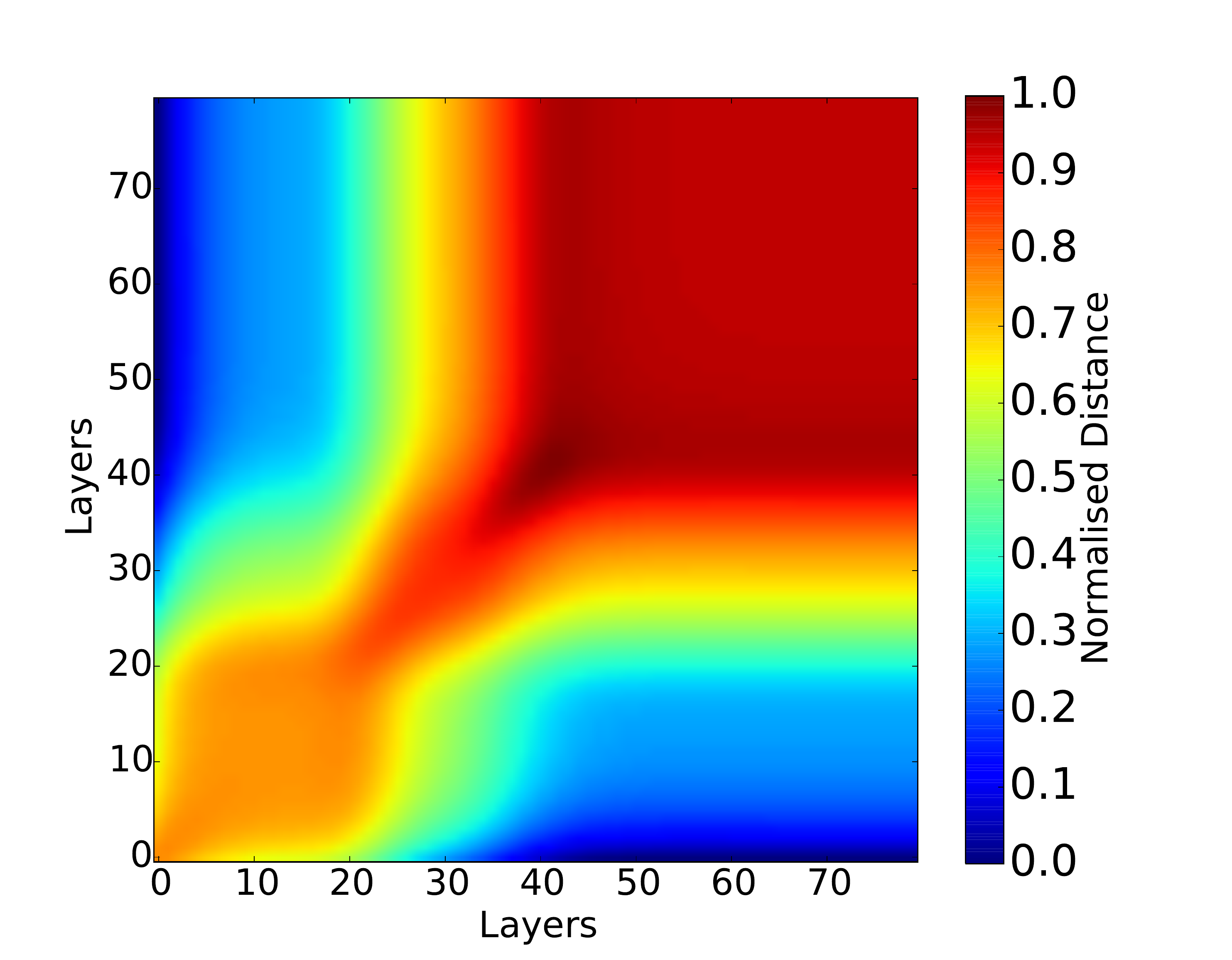}
\caption{Correlation matrix, $\mathcal{C}$, derived from {\it Stage~1}
TP-profile shown in figure~\ref{fig:example1_tp}. The input model TP-profile is
shown as black continuous line. Otherwise identical to
figure~\ref{fig:hybridmatrix}.  \label{fig:example1_tpcov}}
\end{figure}

\begin{figure}
\centering
\includegraphics[width=\columnwidth]{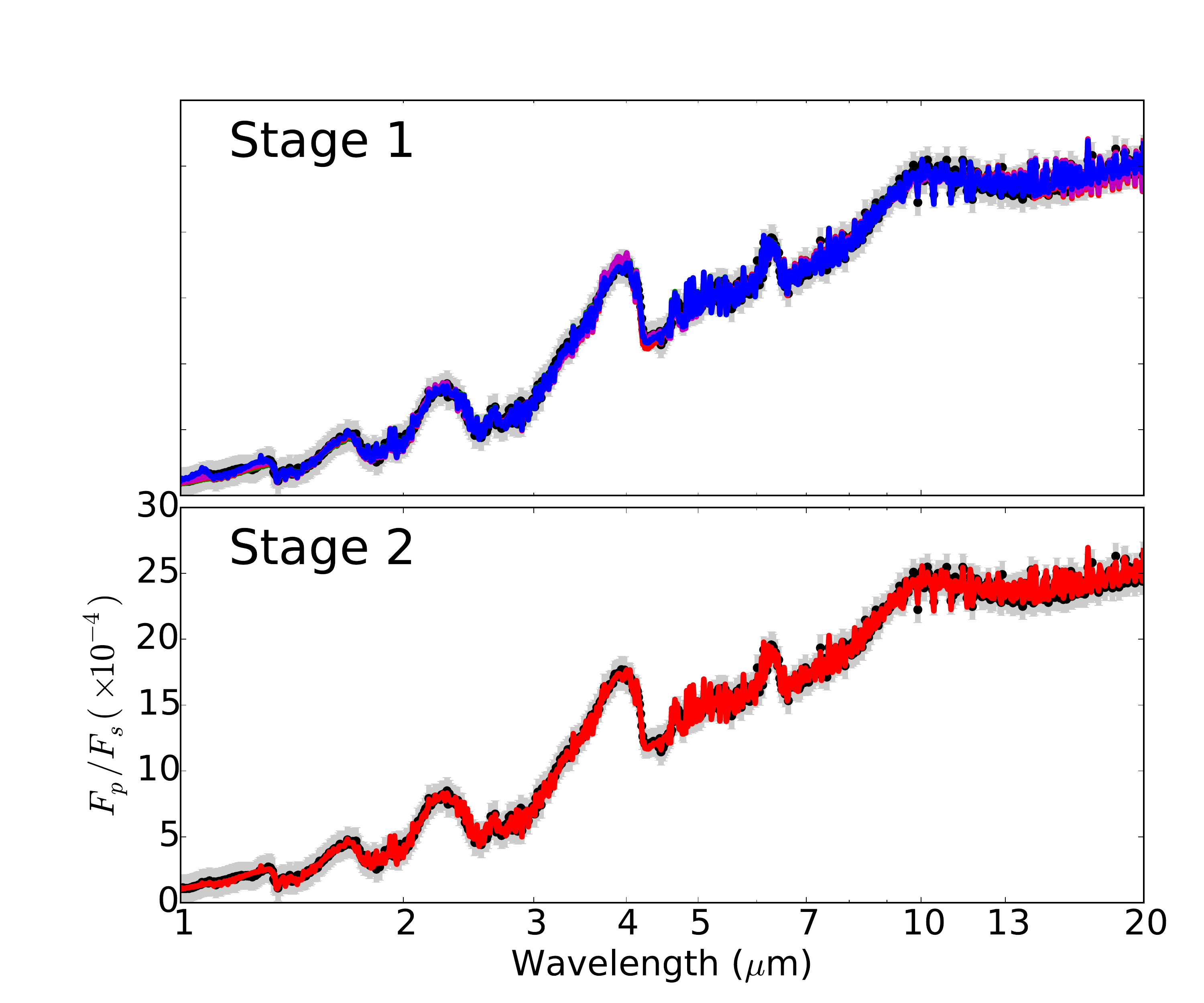}
\caption{showing the input spectrum of a WASP-76b type hot-Jupiter, gray, and
Stage 1~\&~2 fitting on the top and bottom respectively. Both fits converged but
the {\it Stage~2} fit reached a higher maximum likelihood. 
\label{fig:example1_spec}}
\end{figure}

\begin{figure}
\centering
\includegraphics[width=\columnwidth]{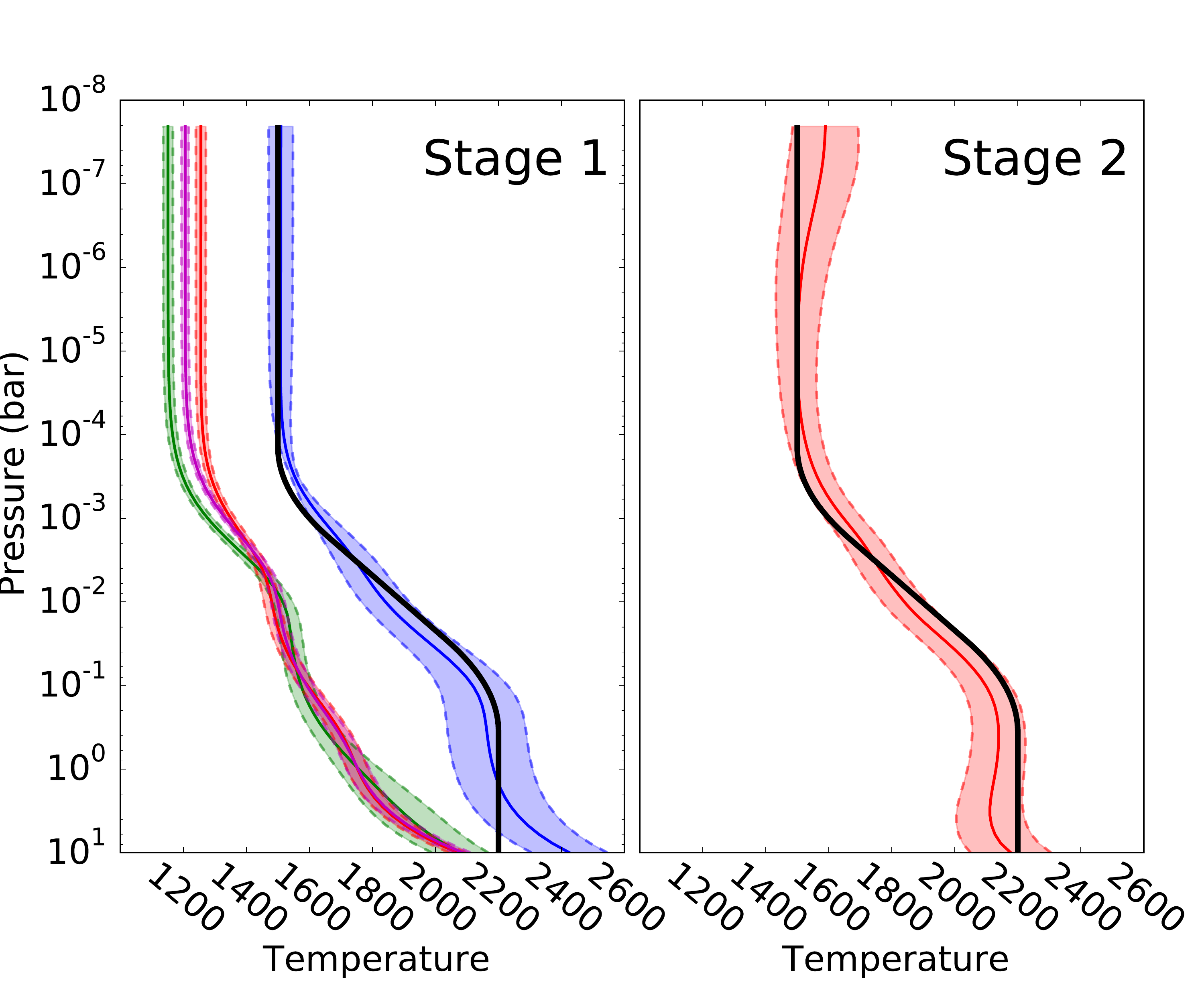}
\caption{TP-profiles for {\it Stage~1}~\&~\emph{2} results
for WASP-76b in figure~\ref{fig:example1_spec}. Solid lines represent the mean and shaded regions the one sigma error bars. Solid black lines are the input TP-profile. {\it Stage~2} takes the initial
parametric TP-profile fit of {\it Stage~1} and relaxes the solution.  Four
solutions were obtained for {\it Stage~1}, the highest maximum likelihood solution (blue) traces the input TP-profile well, whilst local maxima underestimate the bulk temperature, see text. The {\it Stage~2} solution feature a significantly lower
$\chi^{2}$ (or higher global Evidence) than all {\it Stage~1} solutions.
\label{fig:example1_tp}}
\end{figure}

\begin{figure}
\centering
\includegraphics[width=\columnwidth]{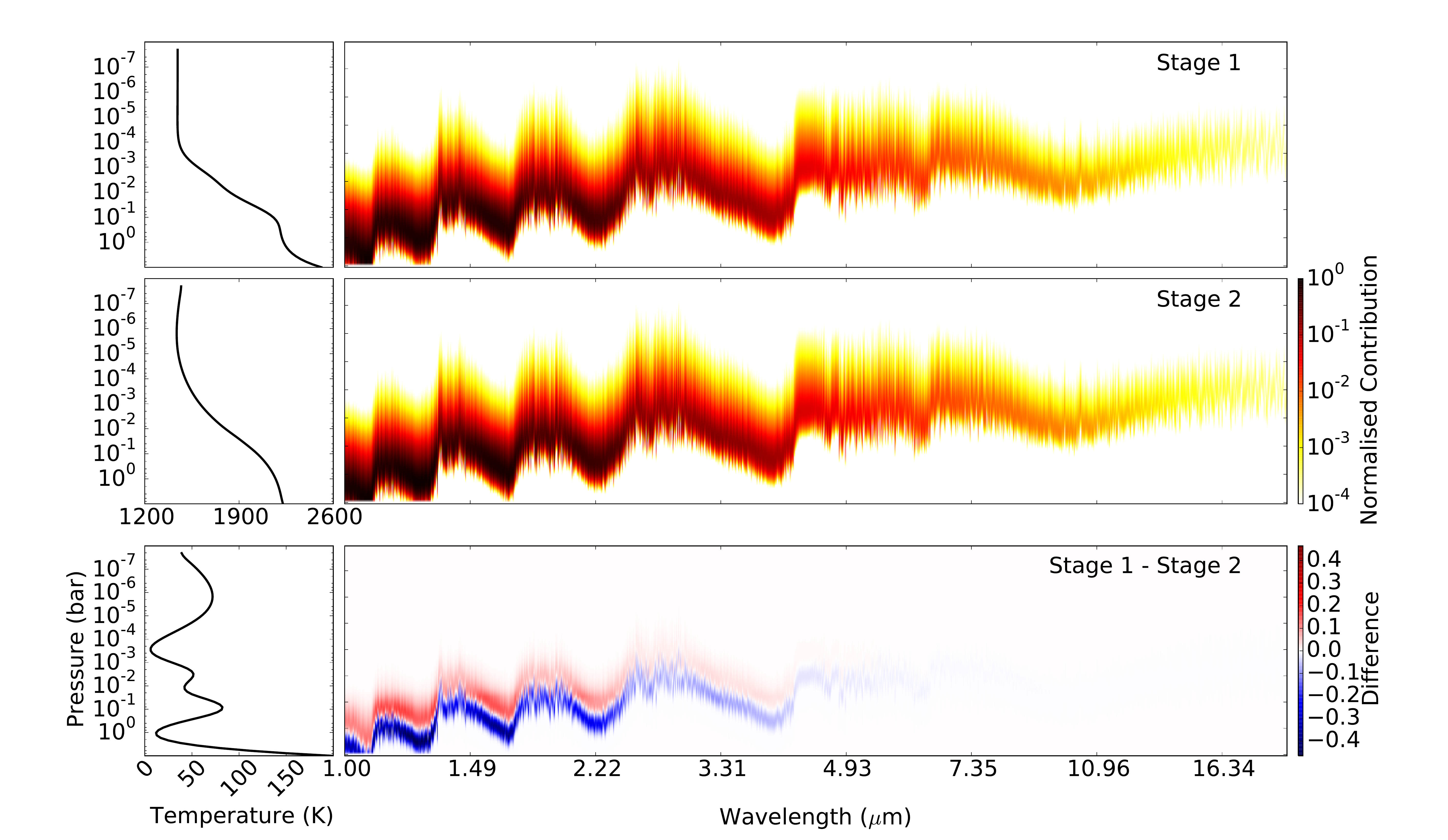}
\caption{Left, retrieved TP-profile, right: contribution functions for WASP-76b. First and second row: Best-fit
TP-profiles and corresponding emission contribution functions as function of
pressure and wavelength for {\it Stage~1} and {\it Stage~2} respectively. Bottom
row: Difference between both stages. Contribution difference is given as
normalised fraction. \label{fig:example1_contrib}}
\end{figure}

\begin{table*}
\centering
 \caption{Retrieved abundances for hot Jupiter WASP-76b.
Top row: log Bayesian Evidence for Stage~1~\&~Stage~2 models. Differences above $| \Delta \text{log}(E) = 5 |$ are very significant. Here  $\Delta \text{log}(E) = + 80.12$ indicating a significantly improved fit in Stage~2.  \label{tb:wasp76}}
\begin{tabular}{r | l | l | l} 
\hline
& Input Model & Stage 1 Retrieval & Stage 2 Retrieval \\ \hline\hline
log($E$) & NA& -43.92 & 36.20 \\\hline
log(H$_{2}$O) & -4.0    & -3.458 $\pm$ 0.104 & -3.660 $\pm$ 0.107 \\
log(CO)           & -4.69  & -4.352 $\pm$ 0.143 & -4.548 $\pm$ 0.113 \\
log(CO$_{2}$) & -6.0    & -5.374 $\pm$ 0.120 & -5.428 $\pm$ 0.114 \\
\hline
 \end{tabular}
\end{table*}

WASP-76b \citep{2013arXiv1310.5607W} is a very hot-Jupiter orbiting a late F7
star. It is highly inflated at $1.83^{+0.06}_{-0.04}$ $R_{Jup}$, $0.92 \pm 0.03$
$M_{Jup}$ and $T_{equ} \sim 2200$K. We take its bulk and orbital properties and
generate a simulated observation at a resolution of 300, spectral range of 0.5 -
20$\mu$m and 100ppm errors. We set the atmospheric composition to $1\times
10^{-4}$ H$_{2}$O, $1 \times 10^{-5}$ CO and $1\times 10^{-5}$ CO$_{2}$. The input TP-profile 
is shown in figure~\ref{fig:example1_tp} (black line). 
It is important to note that here (as well as in section~\ref{sec:55cnc}) the input
TP-profile was generated using a script external to \taurex~with no relation to
either {\it Stage~1} parametrisation. This provides an adequate test-bed for the
{\it Stage~1} fitting to accurately retrieve an arbitrary atmospheric profile.

As described in section~\ref{sec:hybrid}, we retrieve the TP-profile and
abundances in two stages. For computational efficiency reasons, here we only
compute the Nested Sampling solutions (which are also the most accurate, see
W15). Tests were run with both maximum-likelihood retrievals and MCMC retrievals
and solutions between all sets of solutions are in good agreement. 

The Nested Sampling {\it Stage~1} solution returns four local likelihood maxima
of which the global maximum was selected. Figures~\ref{fig:example1_spec}
(top),~\ref{fig:example1_tp} (left) and table~\ref{tb:wasp76} show the retrieved
spectrum, TP-profile and retrieved abundances respectively. The {\it Stage~1}
TP-profile mostly captures the input TP-profile but shows unrealistic bumps and
wiggles as well as unrealistic distributions of the one sigma error bar. These
are artefacts of the parameterisation in section~\ref{sec:parametric}.
The three local maxima TP-profiles shown in figure~\ref{fig:example1_tp} underestimate the bulk temperature of the planet, driving the retrieval to assume unrealistically low abundances of molecular trace gases. In this example this is found to be a numerical effect that disappears by increasing the sampling grid density of the Nested Sampling. Nonetheless the potential degeneracy between TP-profile and molecular abundances is in atmospheric retrievals is worth noting.

As described earlier, we computed the TP-profile covariance, $\mathcal{C}$
(figure~\ref{fig:example1_tpcov}), and tuned the {\it Stage~1} retrieval by
relaxing the parametric solution. Figures~\ref{fig:example1_spec}
(bottom),~\ref{fig:example1_tp} (right) and table~\ref{tb:wasp76} show the {\it
Stage~2} retrieval results. Inspecting figure~\ref{fig:example1_spec}, both {\it
Stage~1} and {\it Stage~2} retrievals fit the input spectrum sufficiently well
but {\it Stage~2} comes significantly closer to the `true' TP-profile and trace
gas abundances. Figure~\ref{fig:example1_contrib} shows the normalised emission
contribution functions of both retrievals and their difference. This shows the
planetary emission mainly emanating in the non-isothermal regions (up to the
tropopause) of the TP-profile, as expected. However {\it Stage~2} emissions
originate significantly lower (by nearly an order of magnitude) in the
atmosphere (blue band in the bottom panel).

\subsection{55 Cnc e}
\label{sec:55cnc}

\begin{figure}
\includegraphics[width=\columnwidth]{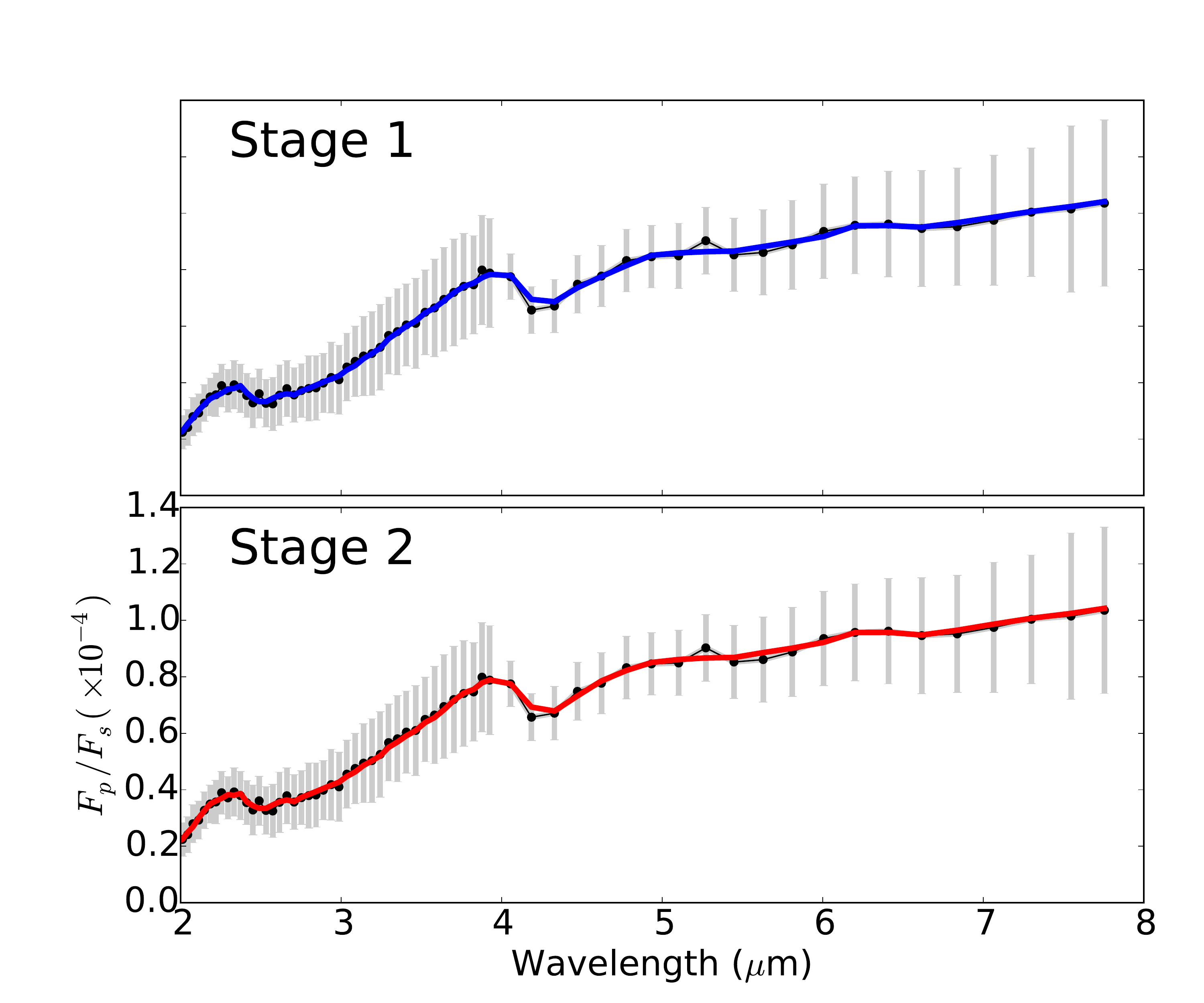}
\caption{Input spectrum of a 55 Cnc e type atmosphere, gray, and Stage 1~\&~2
fitting on the top and bottom respectively. Both fits converged but the {\it
Stage~2} fit reached a higher global Evidence. \label{fig:example2_spec}}
\end{figure}

\begin{figure}
\includegraphics[width=\columnwidth]{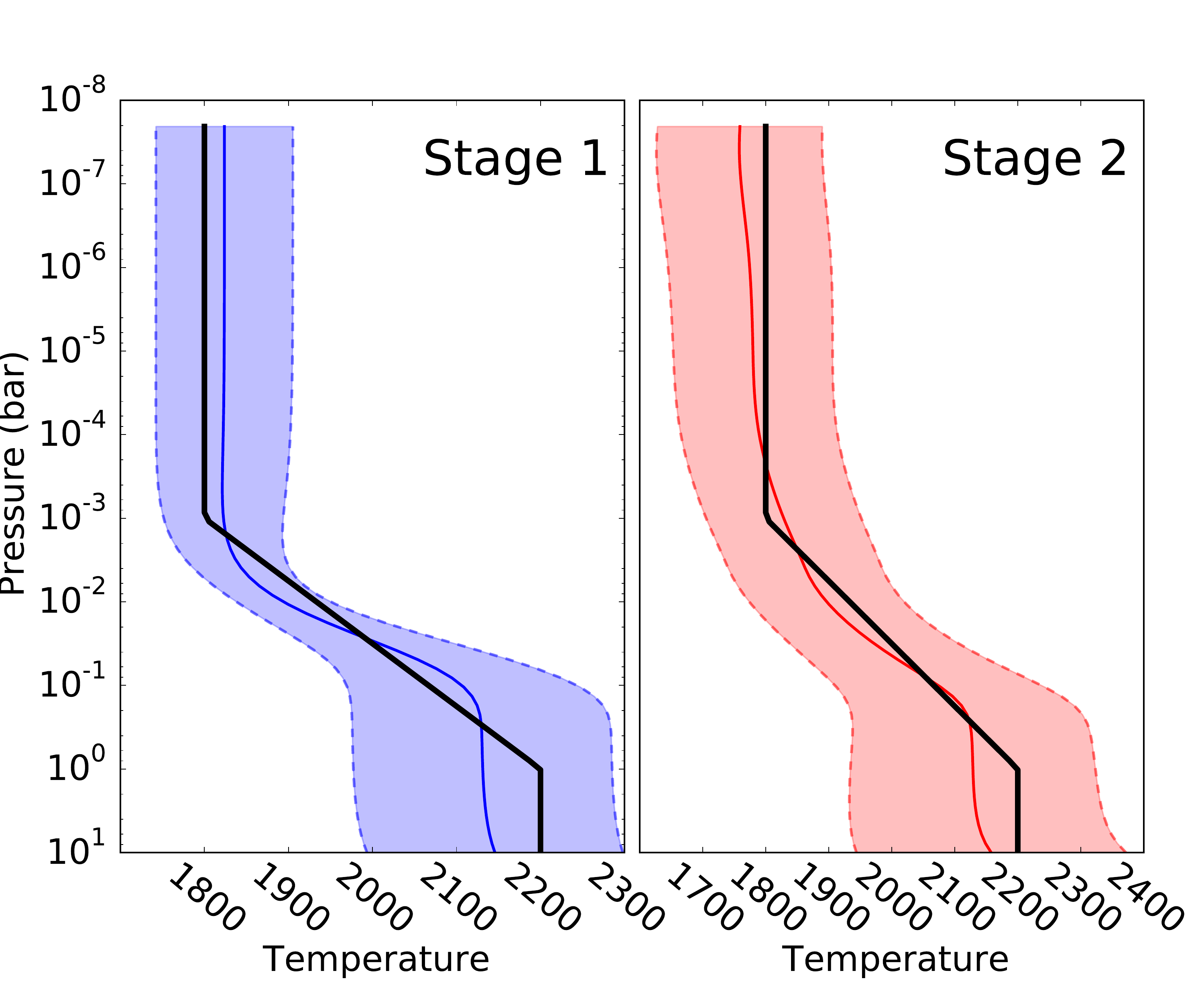}
\caption{TP-profiles for {\it Stage~1}~\&~2 results for 55 Cnc e in
figure~\ref{fig:example2_spec}. Both solutions converge within the calculated
error bar. {\it Stage~2} features a significant improvement in maximum
likelihood achieved.  \label{fig:example2_tp}}
\end{figure}

\begin{figure}
\includegraphics[width=\columnwidth]{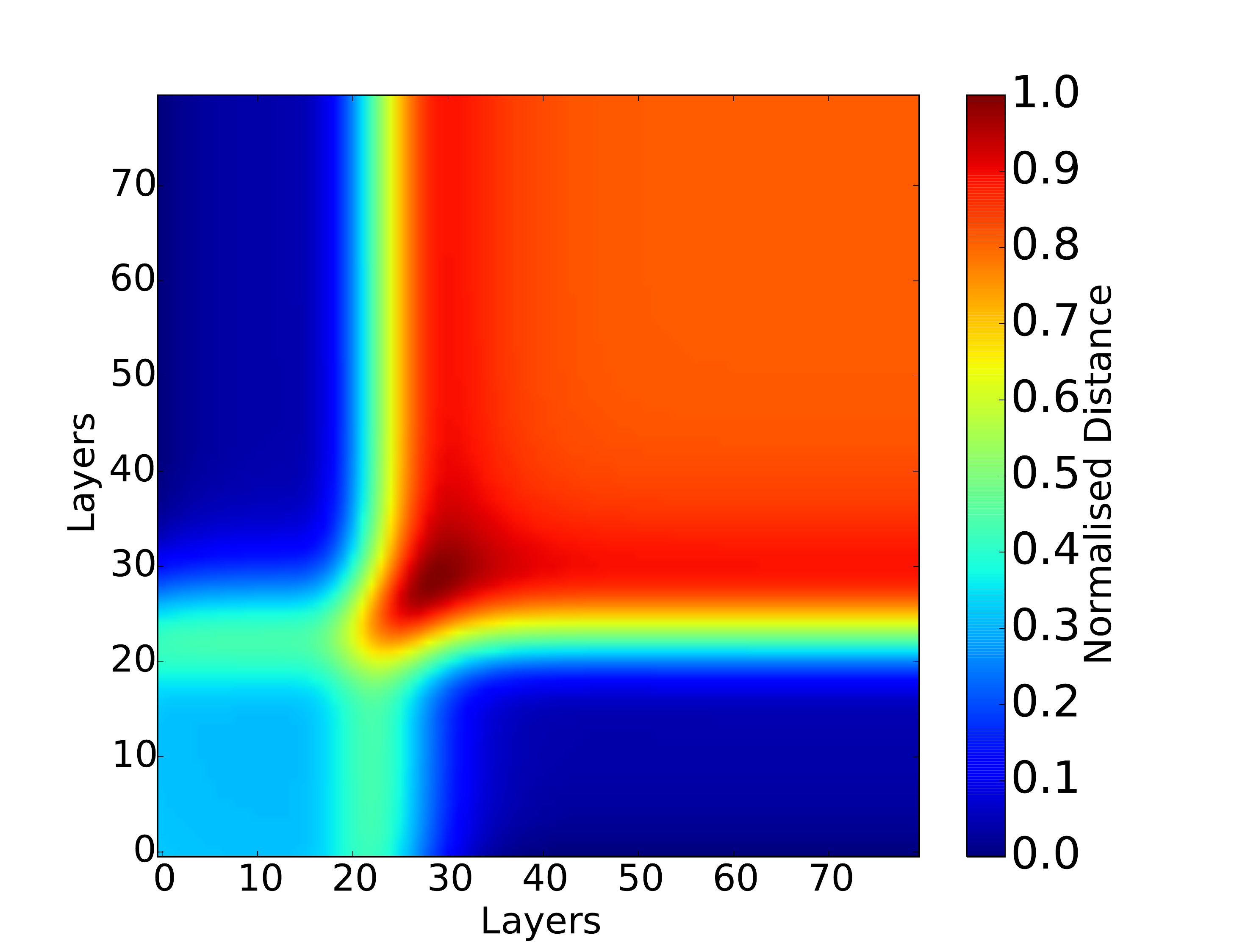}
\caption{Correlation matrix, $\mathcal{C}$, derived from {\it Stage~1}
TP-profile shown in figure~\ref{fig:example2_tp}. Otherwise identical to
figure~\ref{fig:example1_tpcov}.\label{fig:example2_tpcov}}
\end{figure}

\begin{figure}
\includegraphics[width=\columnwidth]{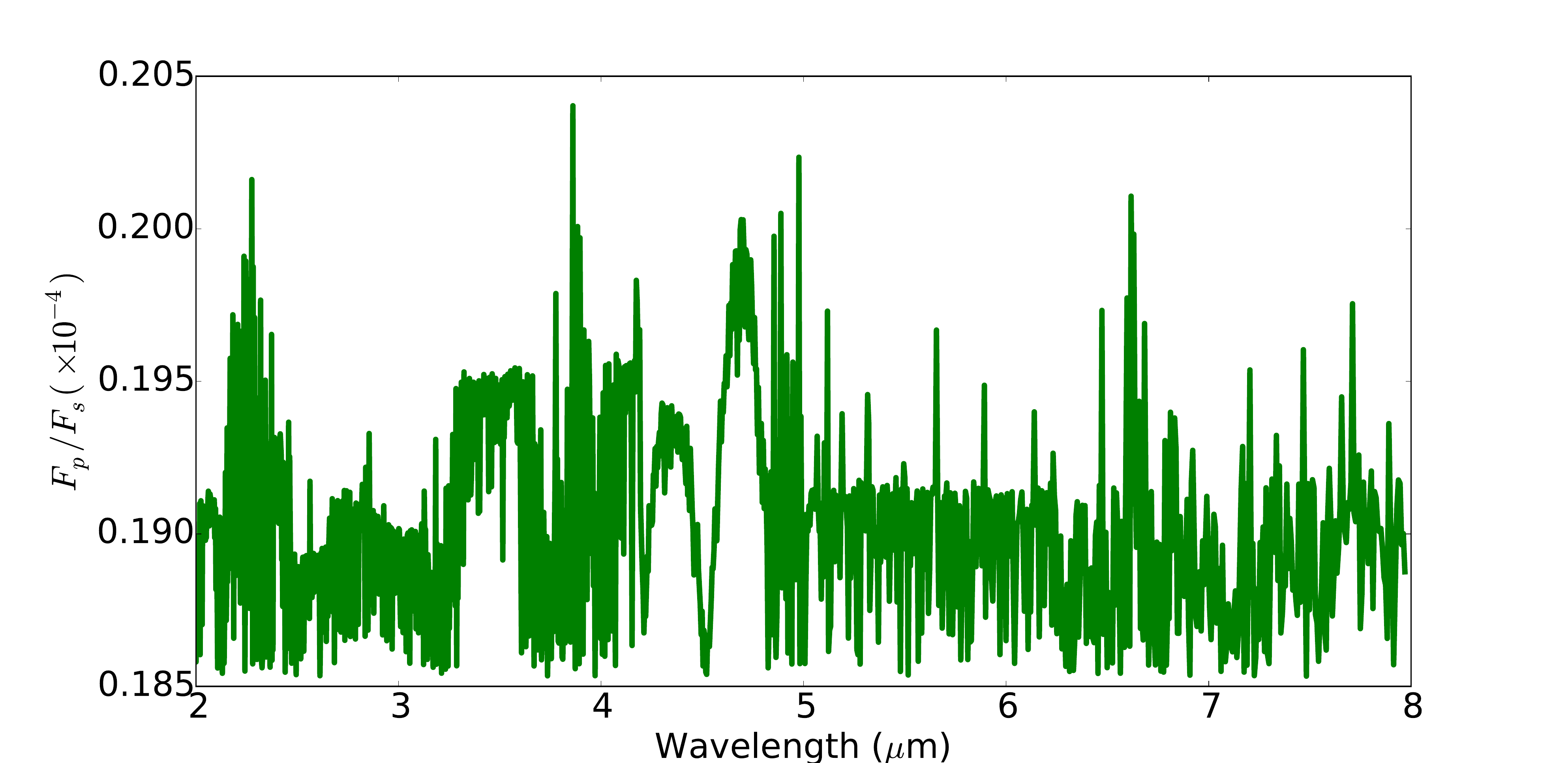}
\caption{showing the model fit difference between Stage 1 and Stage 2 at
spectral resolution of 1000. Model fit difference are of the order of $2 \times
10^{-5}$ or less.  \label{fig:example2_specdiff}}
\end{figure}

\begin{figure}
\includegraphics[width=\columnwidth]{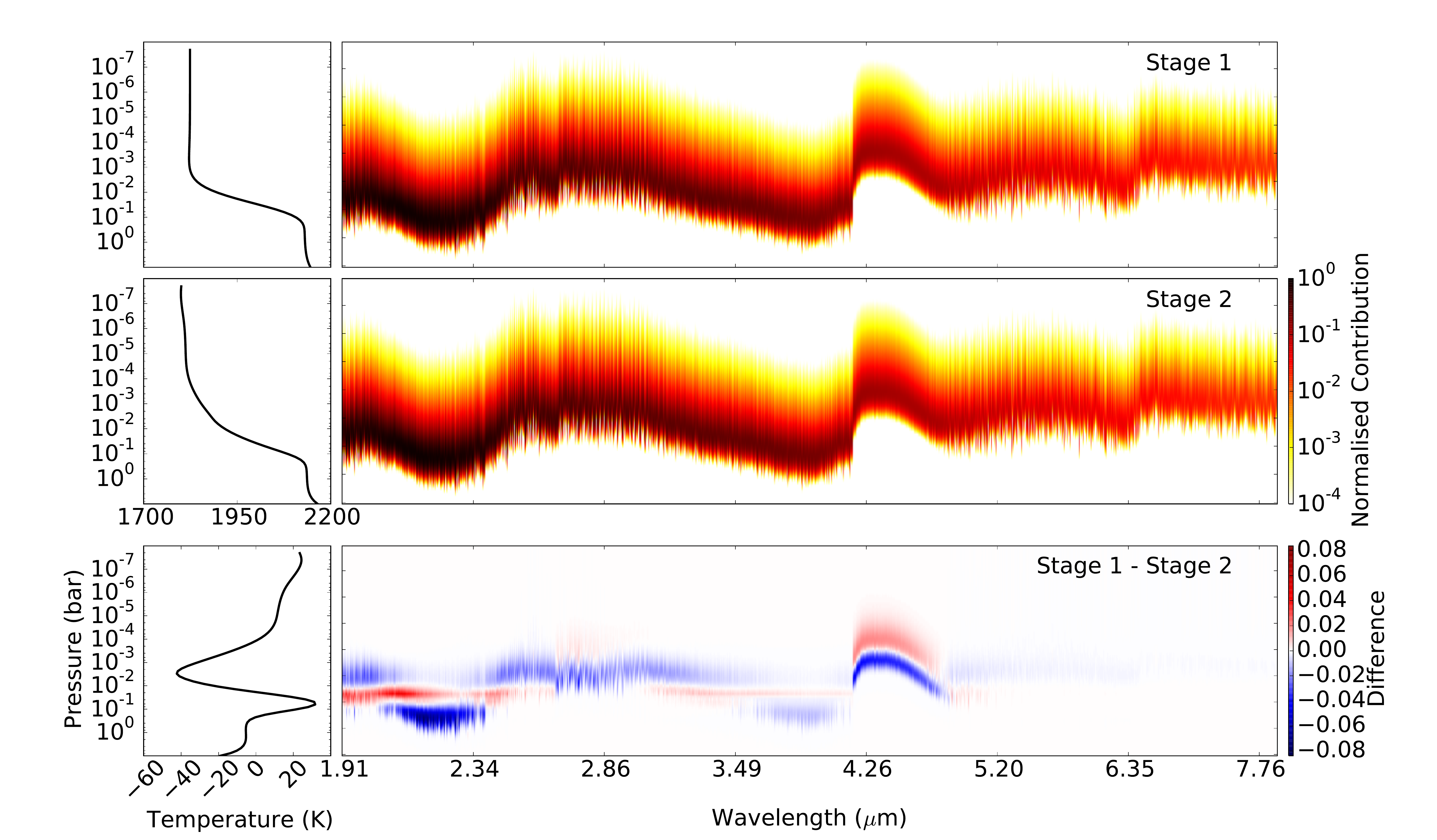}
\caption{Contribution functions for 55Cnc~e, otherwise same as
figure~\ref{fig:example1_contrib}. \label{fig:example2_contrib}}
\end{figure}

\begin{table*}
\centering
\caption{Retrieved abundances for hot SuperEarth 55 Cnc e.
Top row: log Bayesian Evidence for Stage~1~\&~Stage~2 models. \label{tb:55cnc}}
\begin{tabular}{r | l | l | l} 
\hline
& Input Model & Stage 1 Retrieval & Stage 2 Retrieval \\ \hline\hline
log($E$) & NA& 75.40 & 168.90\\\hline
log(H$_{2}$O) & -4.0 & -4.168 $\pm$ 0.795 & -4.055 $\pm$ 0.571 \\
log(CO)   & -5.0 & -5.764 $\pm$ 1.248 & -5.613 $\pm$ 1.172 \\
log(CO$_{2}$) & -5.0 & -5.236 $\pm$ 1.112 & -5.136 $\pm$ 1.019 \\
\hline
 \end{tabular}
\end{table*}

We simulated an emission spectrum of the hot SuperEarth 55 Cnc e
\citep{Fischer:2008kb}. We use trace gas compositions of $1\times 10^{-4}$
H$_{2}$O, $1 \times 10^{-5}$ CO and $1\times 10^{-5}$ CO$_{2}$ and a sharp
TP-profile shown in fig.\ref{fig:example2_tp}. As the previous WASP-76
example is currently unrealistically optimistic, given the combination of high
S/N, moderate resolution (R$\sim$ 300) and a very broad wavelength coverage, we
opt for a more realistic example here. We calculated the expected resolution and
S/N for 100 co-added eclipses obtained by a one-meter-class transiting
spectroscopy space-mission (e.g. the ESA M4 proposal ARIEL) over a wavelength
range spanning 2 to 8$\mu$m. Using {\it EChOSim}, an end-to-end simulator for
transit spectroscopy space-missions \citep{2014arXiv1406.3984P,Waldmann:2014hy}
developed for the ESA M3 EChO proposal \citep{Tinetti:2012hz}, we calculated
realistic error bars for this hot SuperEarth, shown in figure~\ref{fig:example2_spec}.  

Similarly to section~\ref{sec:wasp},
figs~\ref{fig:example2_spec},~\ref{fig:example2_tp}~\&~\ref{fig:example2_tpcov} show the {\it Stage~1}~\&~2 retrieved spectra, TP-profiles
and {\it Stage~1} TP-profile covariance respectively. Table~\ref{tb:55cnc} shows
that {\it Stage~2} retrieval converges at a significantly higher global Evidence
and presents an improvement in the accuracy of abundances retrieved
as well as TP-profile retrieved. Figure~\ref{fig:example2_specdiff} shows the
absolute difference between the {\it Stage~1} and {\it Stage~2} model fits at a
spectral resolution of 1000. Here the discrepancies between both spectral fits
are of the order of $2 \times 10^{-5}$ or less. This is not very significant in
terms of a naive $\chi^{2}$ fit to relatively low S/N data, but does
significantly drive the retrieval of the TP-profile and trace gas abundances.
This is reflected in the Bayesian Evidence measured for {\it Stage~1} and {\it
Stage~2} models. Figure~\ref{fig:example2_contrib} shows the contribution
functions for both retrievals. Here we have positive and negative temperature
differences (see bottom left plot), resulting in a more complex contribution
differential (bottom left plot) than for WASP-76b. It should be noted that this
contribution differential is highly wavelength dependent and illustrates the
sensitivity of varying wavelength ranges on the TP-profile retrievability.  

\clearpage

\subsection{HD189733b}
\label{sec:hd189}

\begin{figure}
\includegraphics[width=\columnwidth]{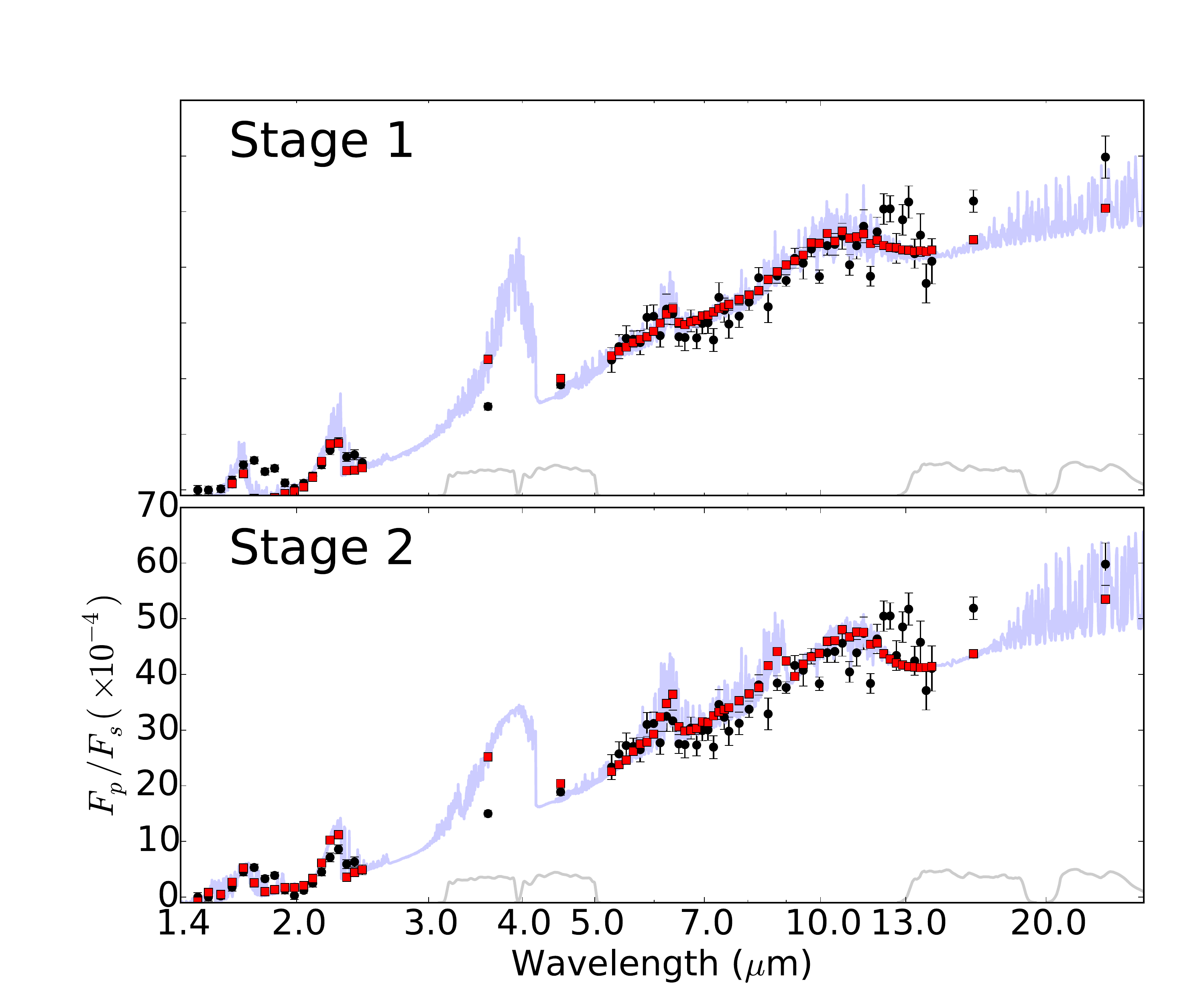}
\caption{Emission spectrum of a HD189733b, black circles, and Stage 1~\&~2
fitting on the top and bottom respectively; blue: the fitted emission spectrum at R=1000; red squares: spectrum fit binned to data resolution; grey: photometric passbands for Spitzer/IRAC and MIPS. 
Both fits converged but the {\it
Stage~2} fit reached a higher global Evidence. \label{fig:hd189_spec}}
\end{figure}

\begin{figure}
\includegraphics[width=\columnwidth]{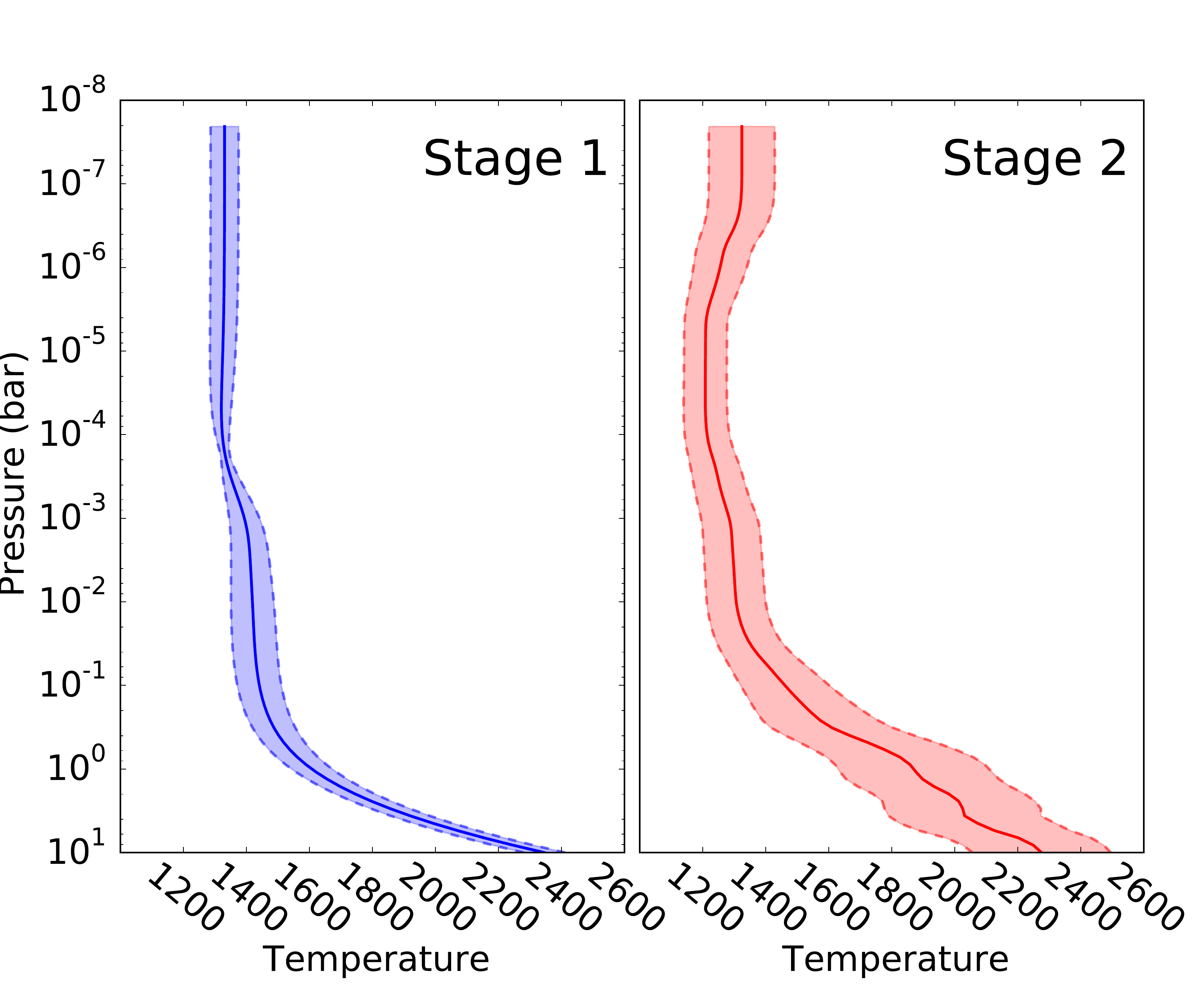}
\caption{TP-profiles for {\it Stage~1}~\&~2 results for HD189733b in
figure~\ref{fig:hd189_spec}. Both solutions converge within the calculated
error bar. {\it Stage~2} features a significant improvement in maximum
likelihood achieved.  \label{fig:hd189_tp}}
\end{figure}

\begin{figure}
\includegraphics[width=\columnwidth]{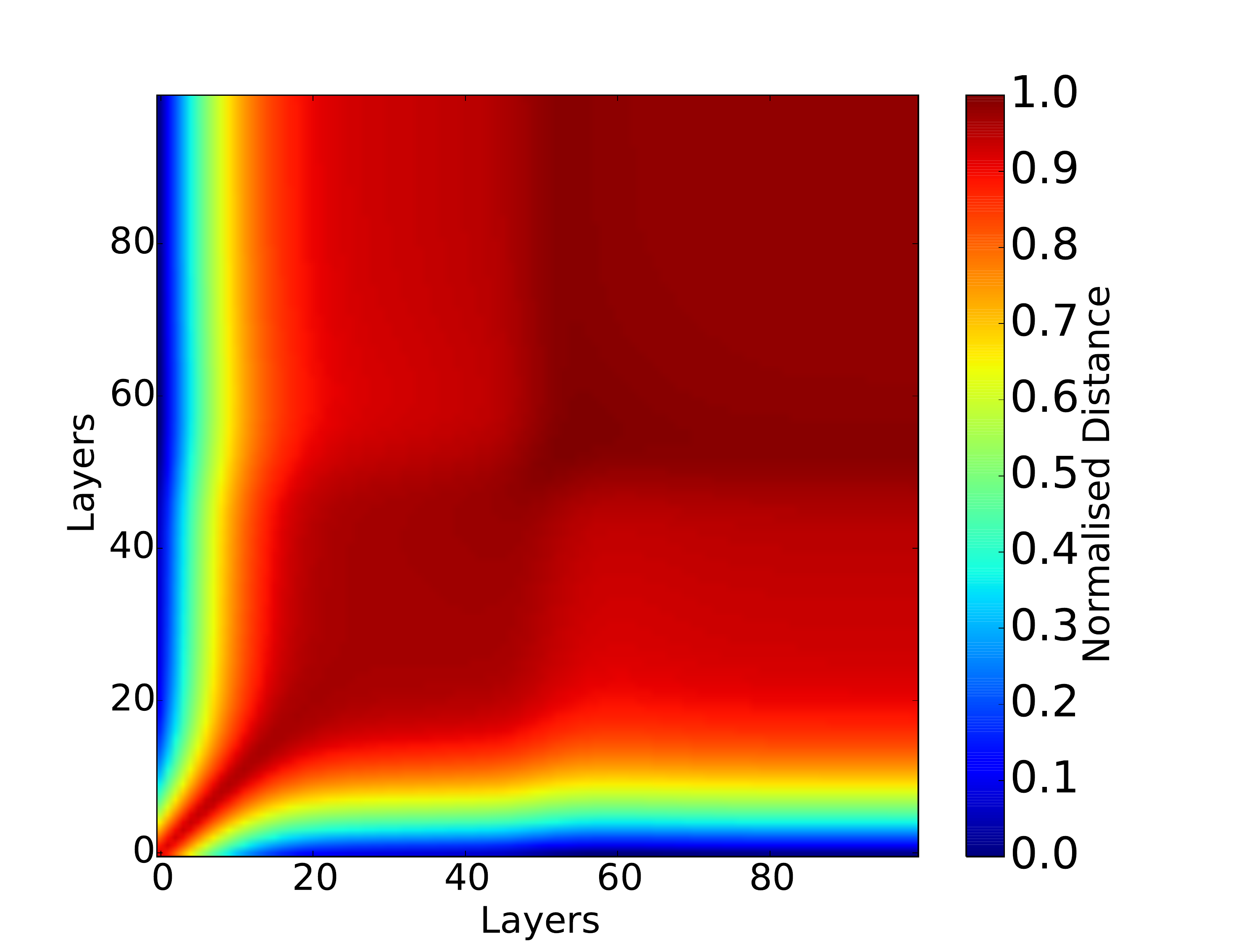}
\caption{Correlation matrix, $\mathcal{C}$, derived from {\it Stage~1}
TP-profile shown in figure~\ref{fig:hd189_tp}. Most of the atmosphere is found to be best fit with an isothermal profile but the lowest $\sim$ 20 atmospheric layers ($\sim$ 0.1 bar). Otherwise identical to
figure~\ref{fig:example1_tpcov}.\label{fig:hd189_tpcov}}
\end{figure}

\begin{figure}
\includegraphics[width=\columnwidth]{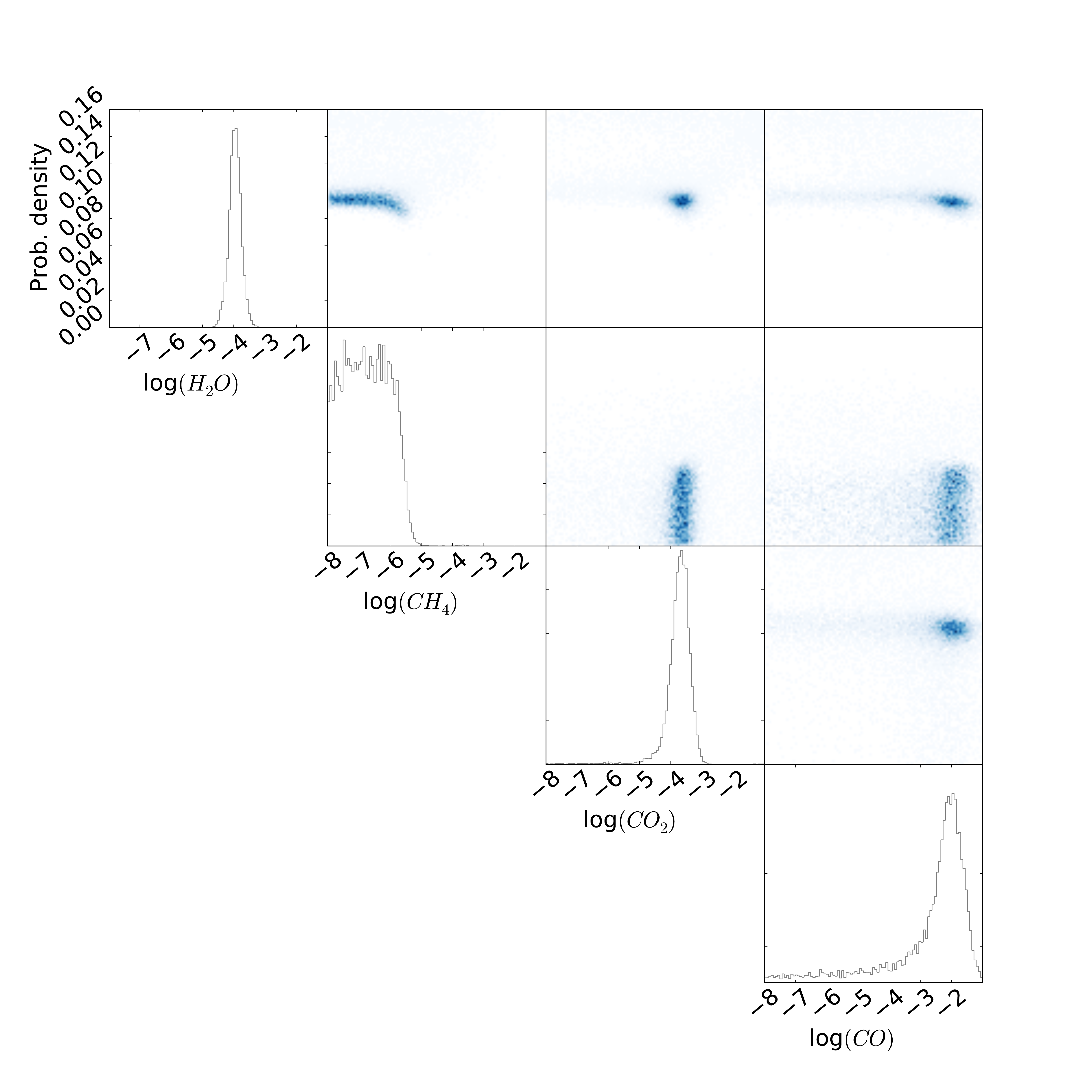}
\caption{Marginalised and conditional posterior distributions of the trace gasses of HD189733b for {\it Stage~1} fitting. All trace gases are well constrained but only an upper limit to methane can be found. \label{fig:hd189_post0}}
\end{figure}

\begin{figure}
\includegraphics[width=\columnwidth]{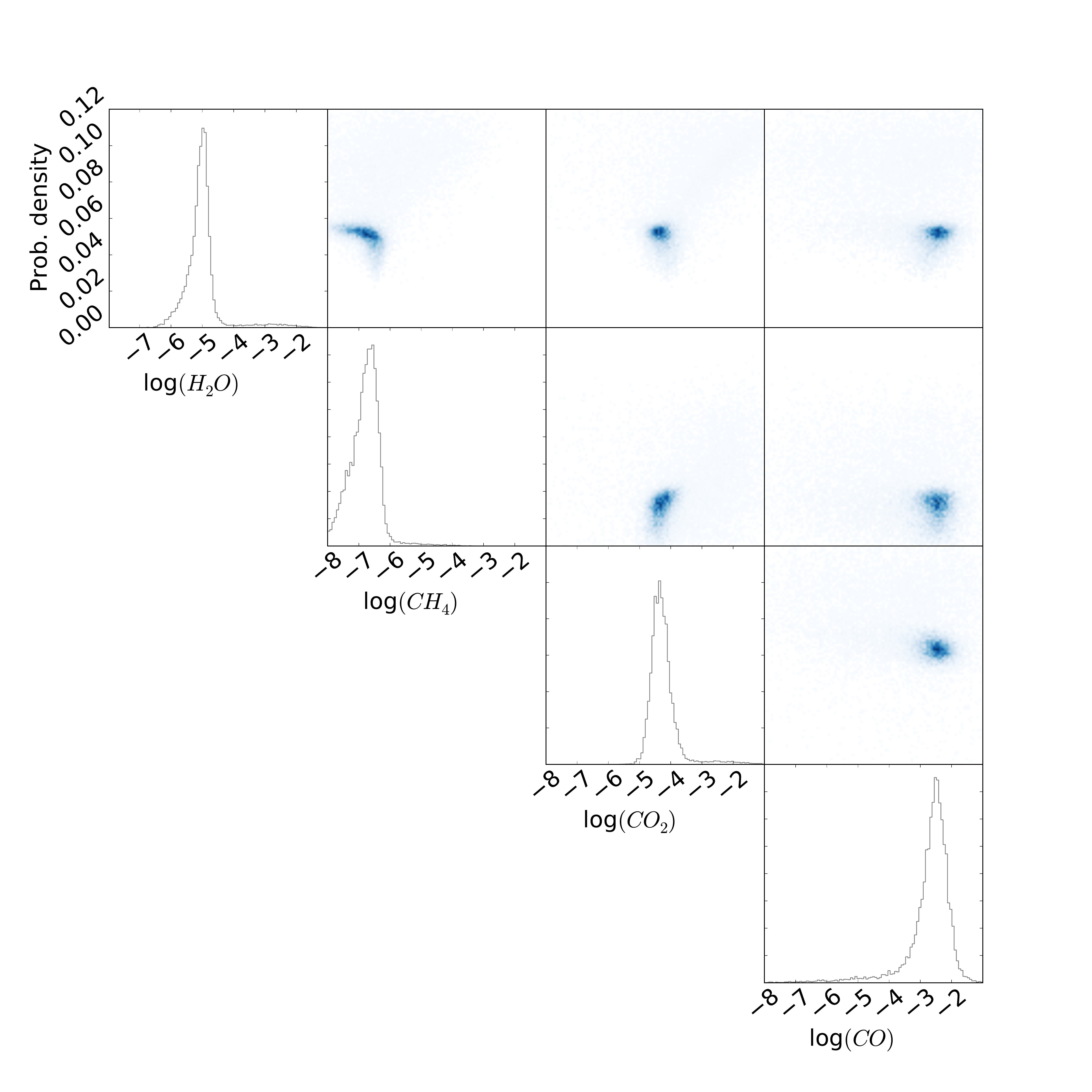}
\caption{Marginalised and conditional posterior distributions of the trace gasses of HD189733b for {\it Stage~2} fitting.  All trace gases are well constrained. \label{fig:hd189_post1}}
\end{figure}

\begin{figure}
\includegraphics[width=\columnwidth]{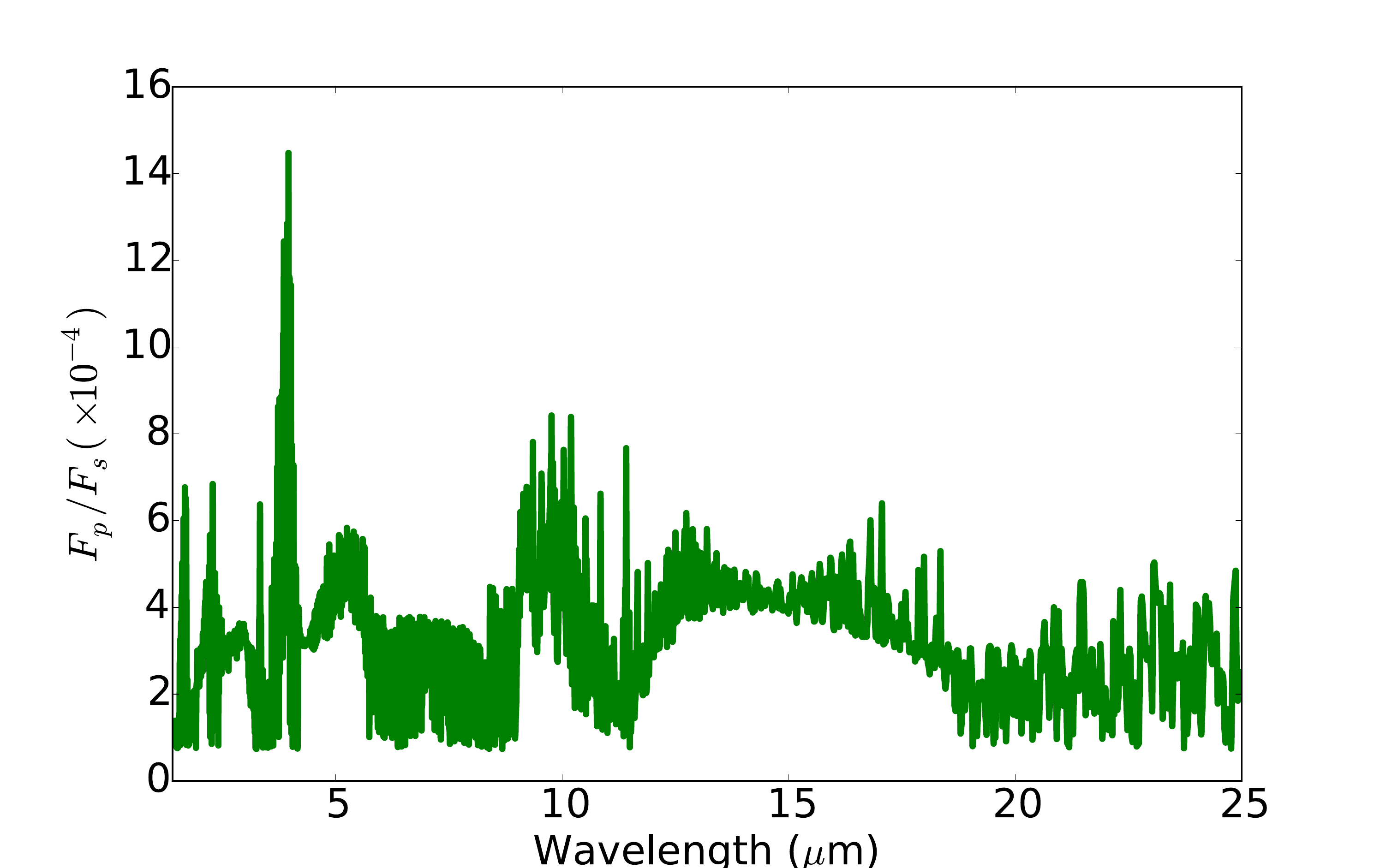}
\caption{Model fit difference between {\it Stage~1} and {\it Stage~2} at
spectral resolution of 1000. Differences between either retrievals are of the order of 5-8$\times10^{-4}$ in most wavelengths. \label{fig:hd189_specdiff}}
\end{figure}

Finally we test \taurex~on the emission spectrum of the hot Jupiter HD189733b. Its emission spectrum is amongst the most complete and best studied \citep[e.g.][]{swain09a,2009ApJ...707...24M,Lee:2011gl,2012ApJ...749...93L,2014ApJ...783...70L}. We have compiled data sets ranging from the NIR (1.4$\mu$m) to the MIR (24$\mu$m), namely:
HST/NICMOS \citep{swain09a}, Spitzer/IRAC 3.6, 4.5 $\mu$m \citep{2012ApJ...754...22K}, 8.0 $\mu$m \citep{2010ApJ...721.1861A}, Spitzer/IRS spectroscopy \citep{2008Natur.456..767G}, Spitzer/IRS 16 \& 24 $\mu$m photometry \citep{2008ApJ...686.1341C}. As in previous studies, we consider the active trace gasses H$_{2}$O, CH$_{4}$, CO and CO$_{2}$ and otherwise assume a hydrogen dominated atmosphere. 

Figure~\ref{fig:hd189_spec} shows the retrieved emission models with retrieved abundances and log(Evidence) reported in table~\ref{tb:hd189}. Figures~\ref{fig:hd189_tp} and \ref{fig:hd189_tpcov} report the retrieved {\it Stage~1} \& {\it Stage~2} TP-profiles and associated {\it Stage~1} TP-covariance, respectively. Figures~\ref{fig:hd189_post0} \& \ref{fig:hd189_post1} are the marginalised and conditional posteriors for the active trace gasses considered. Figure~\ref{fig:hd189_specdiff} illustrates the difference between {\it Stage~1} and {\it Stage~2} emission spectra. 

As described above, the {\it Stage~1} retrieval was first performed using the parmetric TP-profile by \citet{Guillot:2010dd} with the additional optical opacity terms proposed by  \citet{2012ApJ...749...93L}. Here, all parameters of the TP model were allowed to vary. We modelled the atmosphere over 25 scale heights sampled onto a 150 pressure-layer grid.  Following this initial retrieval the TP-covariance was computed resulting in a predominantly isothermal atmosphere down to the lowest $\sim$20 layers ($\sim$ 0.1 bar pressure). The obtained TP-profile is well constrained but shows systematics (e.g. at $5\times10^{-4}$ bar) typical to this type of parametrisation. The posterior distributions of the active trace gases, figure~\ref{fig:hd189_post0} shows good constrains on abundances. From figure~\ref{fig:hd189_post0}, we can see the retrieved values for methane to only constitute an upper limit. 

The {\it Stage~2} hybrid model contained 22 free parameters on its non-linearly sampled TP-profile grid. All parameters were allowed to vary freely between fully constrained ($\alpha = 1.0$ in equation~\ref{equ:hybrid}) and unconstrained scenarios. Figure~\ref{fig:hd189_tp} (right) shows the {\it Stage~2} TP-profile with $\alpha = 0.54$, indicating a significant shift away from the parametric solution of ${\it Stage~1}$. The {\it Stage~2} model features a lower tropopause pressure along with reduced water and carbon-dioxide abundances to achieve a very significant increase in Bayesian Evidence, see table~\ref{tb:hd189} and discussion in section~\ref{sec:discussion}. The {\it Stage~2} TP-profile error bounds are slightly larger. Given a significantly higher Evidence for {\it Stage~2}, it is indicative of {\it Stage~1} TP-profile errors to be underestimated by the parameterisation used. The posteriors distributions of {\it Stage~2} now show a convergence of methane beyond the upper-bound constrained of {\it Stage~1}. 

Compared to table~3 in \citet{2014ApJ...783...70L}, we obtain lower abundances of CH$_{4}$ and CO$_{2}$ but higher values for CO. Whereas these values differ from previous analyses, we find them significantly closer to the chemical model predictions of HD189733b \citep[e.g.][]{line10, venot12, 2013ApJ...763...25M}. Note that also {\it Stage~1} results exhibit the same trend of lower CH$_{4}$ and CO$_{2}$ abundances. We find two aspects by which our analysis differs from others in the literature:
1) A finer and complete sampling of correlated likelihoods, in particular compared to maximum likelihood methods which are often less efficient with sparsely sampled data; 2) The completeness of molecular line-lists used. This is particularly true for the C/O ratio determination using CH$_{4}$ and CO/CO$_{2}$ as tracers. In this work we use the new YT10to10 CH$_{4}$ line-list \citep{2014MNRAS.440.1649Y}.

\begin{table}
\center
\caption{ Retrieved abundances for hot Jupiter HD189733b.
Top row: log Bayesian Evidence for Stage~1~\&~Stage~2 models.   $\Delta \text{log}(E) = + 121.75$ indicating a very significant improvement in the Stage 2 fit compared to Stage 1.  } \label{tb:hd189}
\begin{tabular}{r | l | l } 
\hline
                               & Stage 1 Retrieval                      & Stage 2 Retrieval \\ \hline\hline
log($E$)               & -43.23                         & 78.52\\\hline
log(H$_{2}$O)    & -3.918 $\pm$ 0.207  & -4.978 $\pm$ 0.602 \\
log(CH$_{4}$)    & -6.732 $\pm$ 0.719  & -6.768 $\pm$ 0.487 \\
log(CO$_{2}$)   & -3.722 $\pm$ 0.482   & -4.204 $\pm$ 0.488 \\
log(CO)               & -2.671 $\pm$ 1.387   & -2.689 $\pm$ 0.769 \\
\hline
 \end{tabular}
\end{table}

\section{Discussion}
\label{sec:discussion}

In section~\ref{sec:example} we have demonstrated the efficiency of the \taurex~retrieval suite
for emission spectroscopy using a simulated hot-Jupiter, a hot-SuperEarth, as well as secondary eclipse observations of HD189733b, as examples.

In all three cases the Bayesian Evidence of the {\it Stage~2} retrieval is
significantly higher, log($E$) = 36.20 compared to - 43.92, log($E$) = 168.9 to 75.40, log($E$) = -43.23 to 78.52 for WASP-76b, 55~Cnc~e and HD189733b, respectively.  This is clearly
indicative of {\it Stage~2} results being more robust statistically. Following
the adaptation of  the Jeffrey's scale of model evidence \citep{Jeffreys:1998tt}
by  \citet{Kass:1995vb}, we can define a strong preference for the {\it Stage~2}
model as $\Delta \text{log}(E) > 5$ and equally, a strong preference of the {\it
Stage~1} model to be $\Delta \text{log} (E)< -5$. Evidence differences ranging
from -5 to 5 indicate a lesser significance. In the case of WASP-76b we find
$\Delta \text{log}(E) = + 80.12$, $\Delta \text{log}(E) = + 132.70$  for
55~Cnc~e and $\Delta \text{log}(E) = +121.75$ for HD189733b. Furthermore the improved {\it Stage~2} fit is illustrated by the
better retrieval of the trace gas abundances
(tables~\ref{tb:wasp76},~\ref{tb:55cnc}  \& \ref{tb:hd189}). This illustrates the importance of
accurate TP-profile retrievals and the advantage of a two-staged approach,
especially in cases of low resolution and/or low S/N data. 
 
\section{Conclusion}
In this paper we introduced the emission spectroscopy retrieval approach for the
\taurex~retrieval suite framework. Given a common code basis for transmission
and emission retrieval, allows us to benefit from computational efficiencies and
the high accuracy of molecular line-list handling introduced in W15. To suite the
needs of the temperature-pressure profile retrieval, we implemented an extra
loop unique to the emission side of \taurex, which allows a two-staged
retrieval. We show that such a staged retrieval of the emission spectrum (and
TP-profile) allows us to dynamically scale the complexity of the retrieval
problem (from a fully parameterised to a fully unconstrained model) and has
significant advantages in accuracy and robustness over previously available
methods.  Future publications will extend the sensitivity analyses presented
here to include present and future ground and space instrumentation and a wider
range of planets observable in emission spectroscopy.

 \section*{Acknowledgements}
We thank the referee for providing useful comments. This work was supported by the ERC project numbers 617119 (ExoLights) and
267219 (ExoMol).

%
%
%

\bibliographystyle{apj}
\bibliography{taurex-lib}

\begin{thebibliography}{}
\expandafter\ifx\csname natexlab\endcsname\relax\def\natexlab#1{#1}\fi

\bibitem[{{Agol} {et~al.}(2010){Agol}, {Cowan}, {Knutson}, {Deming}, {Steffen},
  {Henry}, \& {Charbonneau}}]{2010ApJ...721.1861A}
{Agol}, E., {Cowan}, N.~B., {Knutson}, H.~A., {et~al.} 2010, \apj, 721, 1861

\bibitem[{Allard {et~al.}(2012)Allard, Homeier, \& Freytag}]{Allard:2012jx}
Allard, F., Homeier, D., \& Freytag, B. 2012, Philosophical Transactions of the
  Royal Society A: Mathematical, Physical and Engineering Sciences, 370, 2765

\bibitem[{Burrows {et~al.}(2008)Burrows, Budaj, \&
  Hubeny}]{2008ApJ...678.1436B}
Burrows, A., Budaj, J., \& Hubeny, I. 2008, The Astrophysical Journal, 678,
  1436

\bibitem[{Chandrasekar(1960)}]{chandrasekhar1960}
Chandrasekar, H. 1960, {Radiative Transfer} (New York: Dover Publications)

\bibitem[{{Charbonneau} {et~al.}(2008){Charbonneau}, {Knutson}, {Barman},
  {Allen}, {Mayor}, {Megeath}, {Queloz}, \& {Udry}}]{2008ApJ...686.1341C}
{Charbonneau}, D., {Knutson}, H.~A., {Barman}, T., {et~al.} 2008, \apj, 686,
  1341

\bibitem[{Conrath {et~al.}(1973)Conrath, Curran, Hanel, Kunde, Maguire, Pearl,
  Pirraglia, Welker, \& Burke}]{1973JGR....78.4267C}
Conrath, B., Curran, R., Hanel, R., {et~al.} 1973, Journal of Geophysical
  Research, 78, 4267

\bibitem[{Conrath {et~al.}(1970)Conrath, Hanel, Kunde, \&
  Prabhakara}]{Conrath:1970we}
Conrath, B.~J., Hanel, R.~A., Kunde, V.~G., \& Prabhakara, C. 1970, Journal of
  Geophysical Research, 75, 5831

\bibitem[{Fischer {et~al.}(2008)Fischer, Marcy, Butler, Vogt, Laughlin, Henry,
  Abouav, Peek, Wright, Johnson, McCarthy, \& Isaacson}]{Fischer:2008kb}
Fischer, D.~A., Marcy, G.~W., Butler, R.~P., {et~al.} 2008, APJ, 675, 790

\bibitem[{Fletcher {et~al.}(2007)Fletcher, Irwin, Teanby, Orton, Parrish,
  De~Kok, Howett, Calcutt, Bowles, \& Taylor}]{2007Icar..189..457F}
Fletcher, L.~N., Irwin, P. G.~J., Teanby, N.~A., {et~al.} 2007, ICARUS, 189,
  457

\bibitem[{Goody \& Yung(1989)}]{Yung:vp}
Goody, R.~M., \& Yung, Y.~L. 1989, {Atmospheric Radiation} (Oxford: Oxford
  University Press)

\bibitem[{Griffith(2014)}]{2014RSPTA.37230086G}
Griffith, C.~A. 2014, Philosophical Transactions of the Royal Society A:
  Mathematical, 372, 30086

\bibitem[{{Grillmair} {et~al.}(2008){Grillmair}, {Burrows}, {Charbonneau},
  {Armus}, {Stauffer}, {Meadows}, {van Cleve}, {von Braun}, \&
  {Levine}}]{2008Natur.456..767G}
{Grillmair}, C.~J., {Burrows}, A., {Charbonneau}, D., {et~al.} 2008, \nat, 456,
  767

\bibitem[{Guillot(2010)}]{Guillot:2010dd}
Guillot, T. 2010, Astronomy and Astrophysics, 520, A27

\bibitem[{Hanel {et~al.}(1981)Hanel, Conrath, Flasar, Kunde, Maguire, Pearl,
  Pirraglia, Samuelson, Herath, Allison, Cruikshank, Gautier, Gierasch, Horn,
  Koppany, \& Ponnamperuma}]{1981Sci...212..192H}
Hanel, R., Conrath, B., Flasar, F.~M., {et~al.} 1981, Science, 212, 192

\bibitem[{Hanel {et~al.}(2003)Hanel, Conrath, Jennings, \&
  Samuelson}]{Hanel:2003hq}
Hanel, R.~A., Conrath, B., J., Jennings, D.~E., \& Samuelson, R.~E. 2003,
  {Exploration of the Solar System by Infrared Remote Sensing} (Cambridge:
  Cambridge University Press)

\bibitem[{Hanel {et~al.}(1972)Hanel, Conrath, Hovis, Kunde, Lowman, Pearl,
  Prabhakara, Schlachman, \& Levin}]{1972Sci...175..305H}
Hanel, R.~A., Conrath, B.~J., Hovis, W.~A., {et~al.} 1972, Science, 175, 305

\bibitem[{Hansen(2008)}]{2008ApJS..179..484H}
Hansen, B. M.~S. 2008, The Astrophysical Journal Supplement Series, 179, 484

\bibitem[{Heng {et~al.}(2012)Heng, Hayek, Pont, \& Sing}]{2012MNRAS.420...20H}
Heng, K., Hayek, W., Pont, F., \& Sing, D.~K. 2012, Monthly Notices of the
  Royal Astronomical Society, 420, 20

\bibitem[{Heng {et~al.}(2014)Heng, Mendon{\c c}a, \& Lee}]{2014ApJS..215....4H}
Heng, K., Mendon{\c c}a, J.~M., \& Lee, J.-M. 2014, The Astrophysical Journal
  Supplement, 215, 4

\bibitem[{{Hubeny} {et~al.}(2003){Hubeny}, {Burrows}, \&
  {Sudarsky}}]{2003ApJ...594.1011H}
{Hubeny}, I., {Burrows}, A., \& {Sudarsky}, D. 2003, \apj, 594, 1011

\bibitem[{Irwin {et~al.}(2008)Irwin, Teanby, De~Kok, Fletcher, Howett, Tsang,
  Wilson, Calcutt, Nixon, \& Parrish}]{2008JQSRT.109.1136I}
Irwin, P. G.~J., Teanby, N.~A., De~Kok, R., {et~al.} 2008, Journal of
  Quantitative Spectroscopy {\&} Radiative Transfer, 109, 1136

\bibitem[{Jeffreys \& Kendall(1948)}]{Jeffreys:1998tt}
Jeffreys, H., \& Kendall, M.~G. 1948, The Mathematical Gazette, 32, 304

\bibitem[{Kaplan(1959)}]{1959JOSA...49.1004K}
Kaplan, L.~D. 1959, Journal of the Optical Society of America, 49, 1004

\bibitem[{Kass \& Raftery(2012)}]{Kass:1995vb}
Kass, R.~E., \& Raftery, A.~E. 2012, Journal of the American Statistical
  Association, 90, 773

\bibitem[{King(1958)}]{1958sues.conf..133K}
King, J. I.~F. 1958, Scientific Uses of Earth Satellites: Second Revised
  Edition. Edited by James A. Van Allen. Published by the University of
  Michigan Press, -1, 133

\bibitem[{{Knutson} {et~al.}(2012){Knutson}, {Lewis}, {Fortney}, {Burrows},
  {Showman}, {Cowan}, {Agol}, {Aigrain}, {Charbonneau}, {Deming}, {D{\'e}sert},
  {Henry}, {Langton}, \& {Laughlin}}]{2012ApJ...754...22K}
{Knutson}, H.~A., {Lewis}, N., {Fortney}, J.~J., {et~al.} 2012, \apj, 754, 22

\bibitem[{Lee {et~al.}(2011)Lee, Fletcher, \& Irwin}]{Lee:2011gl}
Lee, J.~M., Fletcher, L.~N., \& Irwin, P. G.~J. 2011, Monthly Notices of the
  Royal Astronomical Society, 420, 170

\bibitem[{{Line} {et~al.}(2014){Line}, {Knutson}, {Wolf}, \&
  {Yung}}]{2014ApJ...783...70L}
{Line}, M.~R., {Knutson}, H., {Wolf}, A.~S., \& {Yung}, Y.~L. 2014, \apj, 783,
  70

\bibitem[{Line {et~al.}(2010)Line, Liang, \& Yung}]{line10}
Line, M.~R., Liang, M.~C., \& Yung, Y.~L. 2010, APJ, 717, 496

\bibitem[{Line {et~al.}(2012)Line, Zhang, Vasisht, Natraj, Chen, \&
  Yung}]{2012ApJ...749...93L}
Line, M.~R., Zhang, X., Vasisht, G., {et~al.} 2012, The Astrophysical Journal,
  749, 93

\bibitem[{Line {et~al.}(2013)Line, Wolf, Zhang, Knutson, Kammer, Ellison,
  Deroo, Crisp, \& Yung}]{2013ApJ...775..137L}
Line, M.~R., Wolf, A.~S., Zhang, X., {et~al.} 2013, The Astrophysical Journal,
  775, 137

\bibitem[{Liou(2002)}]{Liou:2002uh}
Liou, K.~N. 2002, {An introduction to atmospheric radiation} (London: Academic
  Press)

\bibitem[{Madhusudhan \& Seager(2009)}]{2009ApJ...707...24M}
Madhusudhan, N., \& Seager, S. 2009, The Astrophysical Journal, 707, 24

\bibitem[{Mihalas \& Mihalas(2013)}]{Mihalas:2013vm}
Mihalas, D., \& Mihalas, B.~W. 2013, {Foundations of Radiation Hydrodynamics}
  (Courier Corporation)

\bibitem[{{Moses} {et~al.}(2013){Moses}, {Madhusudhan}, {Visscher}, \&
  {Freedman}}]{2013ApJ...763...25M}
{Moses}, J.~I., {Madhusudhan}, N., {Visscher}, C., \& {Freedman}, R.~S. 2013,
  \apj, 763, 25

\bibitem[{Parmentier \& Guillot(2014)}]{2014A&A...562A.133P}
Parmentier, V., \& Guillot, T. 2014, Astronomy and Astrophysics, 562, 133

\bibitem[{Pascale {et~al.}(2014)Pascale, Waldmann, MacTavish, Papageorgiou,
  Amaral-Rogers, Varley, de~Foresto, Griffin, Ollivier, Sarkar, Spencer,
  Swinyard, Tessenyi, \& Tinetti}]{2014arXiv1406.3984P}
Pascale, E., Waldmann, I.~P., MacTavish, C.~J., {et~al.} 2014, arXiv.org,
  1406.3984

\bibitem[{Pierrehumbert(2010)}]{2010ppc..book.....P}
Pierrehumbert, R.~T. 2010, Principles of Planetary Climate

\bibitem[{Robinson \& Catling(2012)}]{Robinson:2012ky}
Robinson, T.~D., \& Catling, D.~C. 2012, APJ, 757, 104

\bibitem[{Rodgers(1976)}]{1976RvGSP..14..609R}
Rodgers, C.~D. 1976, Reviews of Geophysics and Space Physics, 14, 609

\bibitem[{Rodgers(2000)}]{Rodgers:2000tw}
---. 2000, {Inverse Methods for Atmospheric Sounding}, Theory and Practice
  (World Scientific Publishing Company Incorporated)

\bibitem[{Rothman {et~al.}(2010)Rothman, Gordon, Barber, Dothe, Gamache,
  Goldman, Perevalov, Tashkun, \& Tennyson}]{2010JQSRT.111.2139R}
Rothman, L.~S., Gordon, I.~E., Barber, R.~J., {et~al.} 2010, Journal of
  Quantitative Spectroscopy {\&} Radiative Transfer, 111, 2139

\bibitem[{Showman {et~al.}(2009)Showman, Fortney, Lian, Marley, Freedman,
  Knutson, \& Charbonneau}]{0004-637X-699-1-564}
Showman, A.~P., Fortney, J.~J., Lian, Y., {et~al.} 2009, The Astrophysical
  Journal, 699, 564

\bibitem[{Swain {et~al.}(2009)Swain, Vasisht, Tinetti, Bouwman, Chen, Yung,
  Deming, \& Deroo}]{swain09a}
Swain, M.~R., Vasisht, G., Tinetti, G., {et~al.} 2009, The Astrophysical
  Journal Letters, 690, L114

\bibitem[{Tennyson \& Yurchenko(2012)}]{Tennyson:2012ca}
Tennyson, J., \& Yurchenko, S.~N. 2012, Monthly Notices of the Royal
  Astronomical Society, 425, 21

\bibitem[{Tinetti {et~al.}(2013)Tinetti, Encrenaz, \&
  Coustenis}]{2013A&ARv..21...63T}
Tinetti, G., Encrenaz, T., \& Coustenis, A. 2013, The Astronomy and
  Astrophysics Review, 21, 63

\bibitem[{Tinetti {et~al.}(2012)Tinetti, Beaulieu, Henning, Meyer, Micela,
  Ribas, Stam, Swain, Krause, Ollivier, Pace, Swinyard, Aylward, Boekel,
  Coradini, Encrenaz, Snellen, Zapatero-Osorio, Bouwman, Cho, Coud{\'e}~de
  Foresto, Guillot, Lopez-Morales, Mueller-Wodarg, Pall{\'e}, Selsis, Sozzetti,
  Ade, Achilleos, Adriani, Agnor, Afonso, Prieto, Bakos, Barber, Barlow,
  Batista, Bernath, B{\'e}zard, Bord{\'e}, Brown, Cassan, Cavarroc, Ciaravella,
  Cockell, Coustenis, Danielski, Decin, Kok, Demangeon, Deroo, Doel, Drossart,
  Fletcher, Focardi, Forget, Fossey, Fouqu{\'e}, Frith, Galand, Gaulme,
  Hern{\'a}ndez, Grasset, Grassi, Grenfell, Griffin, Griffith, Gr{\"o}zinger,
  Guedel, Guio, Hainaut, Hargreaves, Hauschildt, Heng, Heyrovsky, Hueso, Irwin,
  Kaltenegger, Kervella, Kipping, Koskinen, Kovacs, Barbera, Lammer, Lellouch,
  Leto, Lopez~Valverde, Lopez-Puertas, Lovis, Maggio, Maillard,
  Maldonado~Prado, Marquette, Martin-Torres, Maxted, Miller, Molinari, Montes,
  Moro-Martin, Moses, Mousis, Nguyen~Tuong, Nelson, Orton, Pantin, Pascale,
  Pezzuto, Pinfield, Poretti, Prinja, Prisinzano, Rees, Reiners, Samuel,
  Sanchez-Lavega, Forcada, Sasselov, Savini, Sicardy, Smith, Stixrude,
  Strazzulla, Tennyson, Tessenyi, Vasisht, Vinatier, Viti, Waldmann, White,
  Widemann, Wordsworth, Yelle, Yung, \& Yurchenko}]{Tinetti:2012hz}
Tinetti, G., Beaulieu, J.~P., Henning, T., {et~al.} 2012, Experimental
  Astronomy, 34, 311

\bibitem[{{Venot} {et~al.}(2012){Venot}, {H{\'e}brard}, {Ag{\'u}ndez},
  {Dobrijevic}, {Selsis}, {Hersant}, {Iro}, \& {Bounaceur}}]{venot12}
{Venot}, O., {H{\'e}brard}, E., {Ag{\'u}ndez}, M., {et~al.} 2012, \aap, 546,
  A43

\bibitem[{{Waldmann} \& {Pascale}(2014)}]{Waldmann:2014hy}
{Waldmann}, I.~P., \& {Pascale}, E. 2014, Experimental Astronomy,
  arXiv:1402.4408

\bibitem[{Waldmann {et~al.}(2015)Waldmann, Tinetti, Rocchetto, Barton,
  Yurchenko, \& Tennyson}]{waldmann15}
Waldmann, I.~P., Tinetti, G., Rocchetto, M., {et~al.} 2015, The Astrophysical
  Journal, 802, 107

\bibitem[{Wark \& Hilleary(1969)}]{1969Sci...165.1256W}
Wark, D.~Q., \& Hilleary, D.~T. 1969, Science, 165, 1256

\bibitem[{West {et~al.}(2013)West, Almenara, Anderson, Bouchy, Brown,
  Collier-Cameron, Deleuil, Delrez, Doyle, Faedi, Fumel, Gillon, Hebrard,
  Hellier, Jehin, Lendl, Maxted, Pepe, Pollacco, Queloz, Segransan, Smalley,
  Smith, Triaud, \& Udry}]{2013arXiv1310.5607W}
West, R.~G., Almenara, J.~M., Anderson, D.~R., {et~al.} 2013, arXiv.org, 5607

\bibitem[{{Yurchenko} \& {Tennyson}(2014)}]{2014MNRAS.440.1649Y}
{Yurchenko}, S.~N., \& {Tennyson}, J. 2014, \mnras, 440, 1649

\end{thebibliography}

\end{document}